 \documentclass[prd,aps,showpacs,preprintnumbers,amsmath,amssymb,nofootinbib,preprintnumbers]{revtex4}
\usepackage{graphicx}
\usepackage{epsf}
\usepackage{amsmath}
\usepackage{amssymb,latexsym}
\usepackage{epstopdf}
\usepackage{bm}
\usepackage{color}
\usepackage{tabularx}
\usepackage{enumitem}
\usepackage{float}
\usepackage{array,booktabs}
\usepackage{footnote}
\usepackage{threeparttable}
\usepackage{graphicx}
\usepackage{ulem}
\usepackage{hyperref}
\usepackage{amssymb,epsf}
\usepackage{latexsym}
\usepackage{epstopdf}
\usepackage{epsfig}
\usepackage{subfigure}

\begin{document}
     \title{Noether gauge symmetry approach\\
     applying for the non-minimally coupled gravity to the Maxwell field

     }
     \author{
        S. Mahmoudi$^{1,2}$\footnote{email address: S.mahmoudi@shirazu.ac.ir},
        S. Hajkhalili$^{1,2}$\footnote{email address:
        S.hajkhalili@gmail.com} and
        S. H. Hendi$^{1,2,3}$\footnote{email address: hendi@shirazu.ac.ir}
        }
     \affiliation{
        $^1$Department of Physics, School of Science, Shiraz University, Shiraz 71454, Iran \\
        $^2$Biruni Observatory, School of Science, Shiraz University, Shiraz 71454, Iran \\
        $^3$Canadian Quantum Research Center 204-3002 32 Ave Vernon, BC V1T 2L7 Canada }

     \begin{abstract}

     Taking the Noether gauge symmetry approach into account, we
     find spherically symmetric static black hole solutions of the
     non-minimal gauge-gravity Lagrangian of the $\mathcal{R}^\beta F^2$ model.
     At first, we consider a system of differential equations for
     the general non-minimal couplings of $Y(\mathcal{R})F^2$ type, and
     then, we regard a particular $\mathcal{R}^\beta F^2$ non-minimal model
     to find the exact black hole solution and analyze its symmetries.
     As the next step, we calculate the thermodynamical quantities of the black hole
     and study its interesting behavior. Besides, we address thermal stability
     and examine the possibility of the van der Waals-like phase transition.
     \end{abstract}

     \maketitle

\section{Introduction}

Undoubtedly, Einstein's theory of gravity is one of the most
beautiful theories of physics. Although Einstein's theory of
gravity has successfully addressed some essential questions, it
still has very important astrophysical and cosmological problems
such as dark matter and dark energy. To overcome these puzzles,
various approaches have been considered to propose alternative
gravitational theories such as modified gravity models. The
efforts to modify Einstein's theory of gravity have recently
increased by studying some interesting models such as $f(R)$
gravity \cite{Nojiri:2010wj,fR}, scalar-tensor gravity
\cite{Fujii:2003pa}, and vector-tensor gravity. Besides, regarding
the electromagnetic nature of some gravitating systems such as
galaxies, galaxy clusters, stars and planets
\cite{Kim1990,Kim1991,Clarke2001,
Giovannini2004,Widrow2002,Grasso2001}, we have to connect the
Maxwell Lagrangian to gravity model for giving an accurate
description of the mentioned systems.

The modification of Einstein's gravity motivates us to think about
the possible generalization of the Einstein-Maxwell theory in
non-minimal form, especially, in regions of spacetime with very
strong gravitation and gauge fields like near the compact
astrophysical objects with a high mass density such as the neutron
stars or black holes. In such cases, the gravitational field
behaves as a nonlinear medium in which the electromagnetic field
propagates, and therefore, there is the possibility that gravity
and electromagnetic are coupled with each other through
non-minimal coupling. The non-minimal couplings, in general, can
affect the Maxwell and the Einstein field equations: Its effect on
the Maxwell equations can result in the magnetization and the
polarization of a specific medium while its impact on the Einstein
equations may lead to important modifications to the spacetime
metric. If such effects exist, it may be possible to explain some
unexpected observations of gravity such as dark matter, dark
energy, and Pioneer anomaly \cite{Nojiri:2006ri, Harko:2010vs,
Bamba:2008ut,Dereli:2011gh}.

The first attempts in the area of non-minimal modifications were
made by introducing non-minimal couplings of the form $RF^2$ to
get more information about the relationship between spacetime
curvature and the conservation of electric charge \cite{nonmin1,
nonmin2}. Later, they were extended to the general couplings in
$R^mF^2$-type couplings to generate the primordial magnetic fields
in the inflation phase of the universe \cite{Lambiase:2008zz}. On
the other hand, the modified $f(R)$ gravity which has been
proposed to explain the late-time acceleration of the universe has
some black hole solutions in the presence of electromagnetic
fields which are not asymptotically flat \cite{Mazharimousavi}. To
have a well-defined asymptotical behavior and to explain the
late-time acceleration and inflation of the universe, the
extensions of $f(R)$ gravity to the models like $f(R)$-Maxwell
\cite{bamba1,bamba2} and $f(R)$-Yang-Mills gravity \cite{bamba3}
were studied. Moreover, general non-minimal $f(R)$-matter
couplings may explain the rotation curves of test particles
gravitating around galaxies by producing the dark matter effect
\cite{harko1,harko2,nojiri}.\par

\color{black}
In most cases, considering the nonlinearity of the modified
theories, direct method of solving the field equations cannot help
us to obtain solutions. In this regard, the symmetry methods like
the Noether symmetry approach can be regarded as a powerful tool
to calculate the solutions of the field equations. The Noether
symmetry approach which is outlined in \cite{Basilakos:2011rx,
Capozziello:1996bi} is based on the fact that there is a conserved
quantity for every corresponding continuous symmetry
\cite{ibrahim}. Making use of the Noether symmetry approach, it is
possible to obtain conserved quantities concerning the symmetries
of the Lagrangian. The importance of the conserved quantities lies
in the fact that they describe the physical features of
differential equations possessing a Lagrangian in terms of
conservation laws admitted by them. Therefore, with the
conservation relation in hand, the exact solutions of the field
equations can be found. The key role of the Noether symmetry as a
geometric criterion to constrain alternative theories of gravity
has been recently received increasing attentions in cosmological
and astrophysical scales \cite{Dialektopoulos:2018qoe}. In this
regard, the Noether symmetry approach has been applied to study
various cosmological scenarios including scalar-tensor cosmologies
\cite{Dimakis:2017zdu,Dimakis:2017kwx,Giacomini:2017yuk,Paliathanasis:2014rja},
$f(R)$ theories \cite{Capozziello:2008ch,Paliathanasis:2011jq},
nonlocal $f(T)$ gravity \cite{Channuie:2017txg, Nurbaki:2020dgw,
Jamil:2012fs}, $f(R,T)$ theories \cite{Momeni:2015gka} and the
theory of $f(G)$ \cite{Bajardi:2020osh}. Moreover, the authors in
\cite{cosmology1, cosmology2} have studied the exact solutions for
potential functions, scalar field and the scale factors in the
Bianchi models. Apart from the context of cosmology, this approach
has also been useful to find exact solutions in spherically and
axially symmetric spacetime
\cite{Paliathanasis:2011jq,Capozziello:2009jg,Capozziello:2012iea,Paliathanasis:2014iva,Capozziello:2007wc}.

It is also worth noticing that the Noether symmetry can be
considered in two ways: first, the Noether symmetry approach
without a gauge term called {\it{strict Noether symmetry}} that is
a kind of symmetry in which $\mathcal{L}_{X}=0$, where
$\mathcal{L}_{X}$ indicates the Lie derivative of the Lagrangian
arises from the metric of interest along a vector field $X$
\cite{camci1, capo08, vakili08, camci2,darabi2013}. On the other
hand, the Noether symmetry approach with a gauge term known as
{\it{Noether gauge symmetry}} is the generalization of strict
Noether symmetry in the sense that the Noether symmetry equation
includes a divergence of a functional boundary term referred to
the gauge function
\cite{tp2011a,camci2012,yusuf2015,camci2016,Bahamonde:2016}. Both
approaches are useful in a variety of physical problems as well as
applied mathematical ones. However, the key point is that which
approach leads to a conserved quantity. The advantage of Noether
gauge symmetries is that it directly gives conserved quantities or
conservation laws \cite{ibrahim}. This is while the strict Noether
symmetry approach expresses the relationship between Noether's
theorem and cyclic variable. Although the conserved quantities are
also related to the existence of cyclic variables into the
dynamics by the strict Noether symmetry, there is an ambiguity in
the choice of cyclic variables and it is usually required a clever
choice of them. Therefore, the equations for the change of
coordinates have not a unique solution.

Recently, concerning the non-minimal $f(R)$-matter couplings and
in the context of strict Noether symmetry, the general couplings
in $Y(\mathcal{R})F^2$ form were discussed in \cite{Sert:2020vmq}.
Here, we use the Noether gauge symmetry approach to study the
existence of Noether symmetries of gravitational theories which
involved non-minimal coupling with electromagnetic field. We try
to determine possible symmetries as well as corresponding
conserved quantities and evaluate exact solutions for a particular
non-minimal model, i.e. $\mathcal{R}^\beta F^2$.

This paper is organized as follows: In the next section, we
consider the general non-minimal coupling between electromagnetic
field and modified gravity in the form of $Y(\mathcal{R})F^2$ and
try to obtain the system of differential equations using Noether
gauge symmetry approach. In Sec. \ref{3}, we extract the exact
black hole solution and its symmetries and investigate the
geometrical properties of the obtained black hole solution.
Section \ref{4} is devoted to study the thermodynamical properties
of the solution. We calculate the thermodynamic quantities and
examine the first law of thermodynamics. We also explore the
stability and thermal phase transition of the solutions. In the
conclusion part, we give a summary and some concluding remarks.
Finally, in Appendix A, we provide a brief review of the Noether
gauge symmetry approach as complementary information.


\section{System of Differential Equations Coming from Noether Symmetry Condition for the non-minimal
model}\label{2}

In this section, we consider the non-minimal coupling between
electromagnetic field and gravity in the form of
$Y(\mathcal{R})F^2$ which may arise in the presence of a medium
with very high density electromagnetic field and their effects can
be significant even far from the source \cite{Dereli:2011gh}. The
action for the mentioned non-minimal model can be expressed as
\begin{equation}
\mathcal{S} =\,\int d^4x
\sqrt{-g}\,\left(\frac{\mathcal{R}}{2\kappa^2}-\,Y(\mathcal{R})\,F^2\,\right),
\end{equation}\label{action}
where $\kappa^2$ is the universal Newtonian coupling constant of
gravity and $F\equiv F_{\mu\nu} F^{\mu\nu}$ in which $F_{\mu\nu}$
is the Faraday tensor which is constructed using the $U(1)$ gauge
field $A_{\nu}$ as $F_{\mu\nu} =2
\partial_{[\mu}A_{\nu]}$. Regarding the spherically symmetric
static metric
\begin{eqnarray}\label{metric}
ds^2  = - f(r)dt^2 + \frac{1}{f(r)} dr^2 + B(r) \left(d\theta^2+
\sin^2\theta d\varphi^2\right) \ ,
\end{eqnarray}
the corresponding Ricci curvature scalar reads
\begin{eqnarray}\label{Ricci}
\mathcal{R}=-f''-\frac{2fB''}{B}- \frac{2f'B'}{B} +\frac{fB'^2}{
2B^2 } +\frac{2}{B},
\end{eqnarray}
where the prime ($'$) represents the derivative with respect to $r$.\par

Here, we use the gauge Noether symmetry approach given in Appendix
\ref{appendix} to find an exact solution. To this end, canonical
Lagrangian plays an important role. To construct the canonical
Lagrangian for the considered metric \ref{metric}, we choose the
suitable Lagrange multiplier and after integrating by parts, the
Lagrangian $\mathcal{L}$ becomes canonical. Thus the point-like
canonical Lagrangian is derived as follows \cite{Sert:2020vmq}
\begin{eqnarray}\label{PLLag}
\mathcal{L} =\frac{fB'^2}{2 \kappa^2 B} +\frac{f'
B'}{\kappa^2}-\frac{B\mathcal{R}}{2 \kappa^2}+BY(\mathcal{R})\phi
'^2+\frac{2}{\kappa^2},
\end{eqnarray}
where we have assumed that the electromagnetic tensor has only the
electric potential $\phi(r)$, i.e. $A_{\mu}(r)=(\phi(r),0,0,0)$.

The configuration space of \eqref{PLLag} is $\mathcal{Q}$ which
has the generalized coordinates $q^i \equiv \{ f, B, \phi,
\mathcal{R} \}$  and its tangent space $TQ \equiv \{ q^i, q'^i\}
$. Using the Euler-Lagrange equations
\begin{eqnarray}
    \frac{d}{dr} (\frac{\partial \mathcal{L}}{\partial q'^i})-\frac{\partial \mathcal{L}}{\partial q^i}=0
\end{eqnarray}
for the Lagrangian \eqref{PLLag}, we arrive at the following
differential equations for $f$, $B$, $\phi$  and $R$,
respectively,
\begin{eqnarray}\label{difeq1}
&& B''-\frac{B'^2}{2 B}=0   , \\
&& \frac{f'B'}{B}+\frac{fB''}{B}-\frac{fB'^2}{2B^2}+f''+\frac{\mathcal{R}}{2}- Y(\mathcal{R})\phi '^2 \kappa^2=0,    \label{difeq2}
\\
&& [B\phi ' Y(\mathcal{R})]'=0 \label{difeq3},
\\
&& Y_\mathcal{R}(\mathcal{R})\phi '^2=\frac{1}{2 \kappa^2  }  \ . \label{difeq4}
\end{eqnarray}
By inserting the Ricci scalar $\mathcal{R}$ given by \eqref{Ricci}
into the above Eqs. \eqref{difeq1}-\eqref{difeq4}, one can easily
verify that variational field equations are equivalent with those
obtained from the tensorial form of the field equation.\par

Taking into account the variables of the configuration space, the
Noether symmetry generator will be as follows
\begin{equation}\label{eq22}
\mathbf{X} = \xi (r,f,B,\phi,\mathcal{R})\partial_r
+\eta_1\,\partial_f+\,\eta_2\,\partial_B +
\eta_3\,\,\partial_\phi+ \eta_{4}\,\partial_{\mathcal{R}},
\end{equation}
where $\eta_{i}=\eta_i\,(r,f,B,\phi,\mathcal{R})$. Using Eqs.
\eqref{NGSvec}-\eqref{Dif} and imposing the Noether symmetry
condition \ref{condition}
\begin{equation}
X^{[1]} \mathcal{L}+\mathcal{L}(D_t\xi)=D_t \,h,
\end{equation}
for the canonical Lagrangian \eqref{PLLag}, we obtain the
following system of partial differential equations
\begin{eqnarray}\label{eqs}
& & \xi_{,f} = 0, \quad \xi_{,B} = 0, \quad  \xi_{,\mathcal{R}} = 0, \quad \xi_{,\phi} = 0,  \nonumber \\& & \eta_{{1}_{,\mathcal{R}}} = 0, \quad \eta_{{2}_{,f}} = 0, \quad  \eta_{{2}_{,\mathcal{R}}} = 0, \quad \eta_{{3}_{,\mathcal{R}}} = 0, \quad h_{,\mathcal{R}} = 0, \nonumber \\& &
\eta_{1}-\frac{f}{B}\,\eta_{2}\,+\,\frac{2\,B}{\kappa}\,\eta_{{1}_{,B}}\,+\,2\,f\,\eta_{{2}_{,B}}\,-\,f\,\xi_{,r}=0, \nonumber\\&&
B\,Y_{\mathcal{R}}(\mathcal{R})\,\eta_{4}\,+\,B\,Y(\mathcal{R})\,\left(3\,\eta_{{3}_{,\phi}}-\xi_{,r}\,+\,\frac{\eta_{2}}{B}\right)\,=\,0,\nonumber\\&&
\eta_{{1}_{,f}}\,+\,\eta_{{2}_{,B}}\,-\,\xi_{,r}=0,\nonumber\\&&
\eta_{{1}_{,\phi}}\,+\,\kappa\,\frac{f}{B}\,\eta_{{2}_{,\phi}}\,+\,2\,\kappa^{2}\,B\,Y(\mathcal{R})\,\eta_{{3}_{,B}}=0,\nonumber\\&&
\eta_{{2}_{,\phi}}\,+\,2\,B\,\kappa^2\,Y(\mathcal{R})\,\eta_{{3}_{,f}}=0,\nonumber\\&&
\eta_{{2}_{,r}}\,-\,\kappa^2\,h_{,f}\,=0,\nonumber\\&&
\eta_{{1}_{,r}}\,+\,\kappa\,\frac{f}{B}\,\eta_{{2}_{,r}}\,-\,\kappa^2\,h_{,B}\,=\,0,\nonumber\\&&
2\,B\,Y(\mathcal{R})\,\eta_{{3}_{,r}}\,-\,h_{\phi}\,=0,\nonumber\\&&
\left(\frac{2}{\kappa^2}\,-\,\frac{B\,\mathcal{R}}{2\,\kappa^2}\right)\,\xi_{,r}\,-\,\frac{\mathcal{R}}{2\,\kappa^2}\,\eta_{2}\,-\,\frac{B}{2\,\kappa^2}\,\eta_{4}\,-\,h_{,r}\,= 0.
\end{eqnarray}
where $h$ denotes the gauge function (see Appendix A for more
detail). The above system clearly depends on the form of the
function $Y(\mathcal{R})$ and so, by solving it, one can get a
wide class of non-minimal models which are compatible with
spherical symmetry. In the following, we will consider a special
case of $Y(\mathcal{R})$ to search the Noether symmetries.

\section{Non-minimal $\mathcal{R}^{\beta}\,F^{2}$-Coupled Electromagnetic Field to
Gravity}\label{3}

In this section, by choosing a functional form of the non-minimal
coupling between the electromagnetic field and gravity, i.e.
$\mathcal{R}^{\beta}\,F^{2}$, we try to obtain an exact solution
with black hole interpretation. Then, we investigate some
geometrical properties of the black hole solution.

\subsection{The exact solution and its symmetry}
It follows from the Noether symmetry equations given by
\eqref{eqs} that for an arbitrary form of the function
$Y(\mathcal{R})$, the system of partial differential equations
reduces to the following equations
\begin{eqnarray}\label{eqs1}
& & \xi_{,f} = 0, \quad \xi_{,B} = 0, \quad  \xi_{,\mathcal{R}} = 0, \quad \xi_{,\phi} = 0,  \nonumber \\& & \eta_{{1}_{,\mathcal{R}}} = 0, \quad \eta_{{1}_{,\mathcal{\phi}}} = 0,\nonumber \\&& \eta_{{2}_{,f}} = 0, \quad  \eta_{{2}_{,\mathcal{R}}} = 0, \quad \eta_{{2}_{,\phi}} = 0, \nonumber \\&&\eta_{{3}_{,f}} = 0, \quad \eta_{{3}_{,B}} = 0, \quad \eta_{{3}_{,\mathcal{R}}} = 0,  \nonumber \\& & h_{,\mathcal{R}} = 0, \quad h_{\phi}=0, \nonumber \\& &
\eta_{1}-\frac{f}{B}\,\eta_{2}\,+\,\frac{2\,B}{\kappa}\,\eta_{{1}_{,B}}\,+\,2\,f\,\eta_{{2}_{,B}}\,-\,f\,\xi_{,r}=0, \nonumber\\&&
\eta_{{1}_{,f}}\,+\,\eta_{{2}_{,B}}\,-\,\xi_{,r}=0,\nonumber\\&&
3\eta_{{3}_{,\phi}}\,+\,\frac{\eta_{2}}{B}\,-\,\xi_{,r}=0,\nonumber\\&&
\eta_{{2}_{,r}}\,-\,\kappa^2\,h_{,f}\,=0,\nonumber\\&&
\eta_{{1}_{,r}}\,+\,\kappa\,\frac{f}{B}\,\eta_{{2}_{,r}}\,-\,\kappa^2\,h_{,B}\,=\,0,\nonumber\\&&
\left(\frac{2}{\kappa^2}\,-\,\frac{B\,\mathcal{R}}{2\,\kappa^2}\right)\,\xi_{,r}\,-\,\frac{\mathcal{R}}{2\,\kappa^2}\,\eta_{2}\,-\,\frac{B}{2\,\kappa^2}\,\eta_{4}\,-\,h_{,r}\,= 0.
\end{eqnarray}
Solving the above differential equations show that the components of the Noether generator $X$ are
\begin{eqnarray}
&& \xi = \frac{3}{2}c_1 r +c_2, \quad \eta_1 = 3\,c_{1}\,f\,+
\,\left(c_{3}\,r\,+\,c_{4}\right)\,B^{-\frac{\kappa}{2}}, \quad
\eta_2 = -\frac{3}{2}\,c_{1}\,B, \quad \eta_3 =
c_{1}\,\phi\,+\,c_{5}, \quad \eta_4 = 0 , \nonumber\\&&
h\,=\,\frac{3}{\kappa^2}\,c_{1}\,r\,-\,\frac{2}{\kappa^2\,(\kappa-2)}\,c_{3}\,B^{(1-\frac{\kappa}{2})}\,+\,c_{6},
\end{eqnarray}
which yields {\it five} Noether symmetries as
\begin{eqnarray}
& & {\bf X}_1 = \frac{3}{2}\,r\,\partial_r\,+
\,3\,f\,\partial_{f}\,-\,\frac{3}{2}\,B\,\partial_{B}\,+\,\phi\,\partial_{\phi},\nonumber\\&&
 {\bf X}_2 =\,\partial_{r}, \quad\quad \quad  \quad \quad \quad {\bf X}_3 =\,\partial_{\phi},\nonumber\\&&
 {\bf X}_4 =\,r\,B^{-\frac{\kappa}{2}}\,\partial_{f},\quad\quad \quad  {\bf X}_5 =\,B^{-\frac{\kappa}{2}}\,\partial_{f},
\end{eqnarray}
with the following Lie algebra
\begin{eqnarray}
&&\left[{\bf X}_2,{\bf X}_1 \right] =\frac{3}{2}\, {\bf X}_2,
\,\quad\quad \left[{\bf X}_3,{\bf X}_1 \right] = {\bf X}_3,
\quad\quad  \left[{\bf X}_4,{\bf X}_1 \right] =
(\frac{6-3\,\kappa}{4}) {\bf X}_4,\nonumber\\&& \left[{\bf
X}_5,{\bf X}_1 \right] = (3-\frac{3\,\kappa}{4}) {\bf X}_5,
\quad\quad \left[{\bf X}_2,{\bf X}_4 \right] = {\bf X}_5.
\label{commut}
\end{eqnarray}
Then the corresponding first integrals of ${\bf X}_1,{\bf X}_2,
{\bf X}_3$, ${\bf X}_4$ and ${\bf X}_5$   are
\begin{eqnarray}
&& I_1 = \frac{3}{2}\,\frac{f\,B^{'}}{\kappa^2}-\frac{3}{2}\,\frac{B\,f^{'}}{\kappa^2}\,+\,I_{3}\,\phi\,-\,\frac{3}{\kappa^2}\,r\nonumber\nonumber\\
&& I_{2}\,=\,0, \quad\quad\quad\quad \quad  I_{3}\,=\,2\,B\,Y(\mathcal{R})\,\phi^{'}\nonumber\\
&& I_{4}\,=\,r\,I_{5}\,+\,\frac{2}{\kappa^2\,(\kappa-2)}\,B^{(1-\frac{\kappa}{2})}, \nonumber\\
&& I_{5}\,=\,\frac{B^{'}}{\kappa^2}\,B^{-\frac{\kappa}{2}},
\end{eqnarray}
where $I_1, I_2, I_3, I_4$ and $I_5$ are constants of motion. It
is notable that the integration constant $I_3$ corresponds to the
electric charge of the source which is determined by the Gauss
integral
\begin{equation}\label{Gauss}
q=\frac{1}{4\pi}
{\int_{S^2}{\sqrt{-g}}\,Y(\mathcal{R})\,F^{\mu\,\nu}\,dS_{\mu\,\nu}},
\end{equation}
which clearly represents the relation between the electric charge
and the spacetime curvature. It is also possible to find solutions
to the Noether symmetry equations \eqref{eqs} for some specific
functions of $Y(\mathcal{R})$.

Let us consider the function $Y(\mathcal{R})$ in the following form
\begin{equation}\label{Y}
    Y(\mathcal{R})\,=\,\frac{1}{1-(a_{0}\,\mathcal{R})^{\beta}},
\end{equation}
where $a_0$ is a coupling constant with dimension $[L]^2$ and
$\beta$ is a real number. This type of non-minimal couplings
between electromagnetic field and gravity was first proposed by T.
Dereli et. al. in \cite{Dereli:2011gh}. They derived the field
equations by a first order variational principle using the method
of Lagrange multipliers for the spherically symmetric static
metric \cite{Dereli:2011gh}. Here, we also try to find the
solutions from Noether symmetry approach. In other words, we
investigate the possible Noether symmetries of the spherically
symmetric solution for this non-minimal model.

We find from the system \eqref{eqs} that the components of the
Noether generator $X$ for this case are
\begin{eqnarray}
&& \xi =\,c_1 r +c_2, \quad \eta_1 =
2\,c_{1}\,f\,+\,\left(c_{3}\,r\,+\,c_{4}\right)\,B^{-\frac{\kappa}{2}},
\quad \eta_2 = -\,c_{1}\,B, \quad \eta_3 =\, \frac{2}{3}
c_{1}\,\phi\,+\,c_{5}, \quad \eta_4 = 0 , \nonumber\\&&
h\,=\,\frac{2}{\kappa^2}\,c_{1}\,r\,-\,\frac{2}{\kappa^2\,(\kappa-2)}\,c_{3}\,B^{(1-\frac{\kappa}{2})}\,+\,c_{6}.
\end{eqnarray}
Hence, we arrive at the following Noether symmetries
\begin{eqnarray}
& & {\bf X}_1 = \,r\,\partial_r\,+\,2\,f\,\partial_{f}\,-\,\,B\,\partial_{B}\,+\,\frac{2}{3}\,\phi\,\partial_{\phi},\nonumber\\&&
{\bf X}_2 =\,\partial_{r}, \quad\quad \quad  \quad \quad \quad {\bf X}_3 =\,\partial_{\phi},\nonumber\\&&
{\bf X}_4 =\,r\,B^{-\frac{\kappa}{2}}\,\partial_{f},\quad\quad \quad  {\bf X}_5 =\,B^{-\frac{\kappa}{2}}\,\partial_{f},
\end{eqnarray}
which gives rise to the first integrals as
\begin{eqnarray}\label{generator}
&& I_1 = \,\frac{f\,B^{'}}{\kappa^2}-\,\frac{B\,f^{'}}{\kappa^2}\,+\,I_{3}\,\phi\,-\,\frac{2}{\kappa^2}\,r\nonumber\nonumber\\
&& I_{2}\,=\,0, \quad\quad\quad\quad \quad  I_{3}\,=\,2\,B\,Y(\mathcal{R})\,\phi^{'}\nonumber\\
&& I_{4}\,=\,r\,I_{5}\,+\,\frac{2}{\kappa^2\,(\kappa-2)}\,B^{(1-\frac{\kappa}{2})}, \nonumber\\
&& I_{5}\,=\,\frac{B^{'}}{\kappa^2}\,B^{-\frac{\kappa}{2}}.
\end{eqnarray}
Also, the corresponding Lie algebra of Noether symmetries has the
following non-vanishing commutators
\begin{eqnarray}
&&\left[{\bf X}_2,{\bf X}_1 \right] = \, {\bf X}_2, \,\quad\quad
\left[{\bf X}_3,{\bf X}_1 \right] = {\bf X}_3, \quad\quad
\left[{\bf X}_4,{\bf X}_1 \right] = (1-\frac{\kappa}{2}) {\bf
X}_4,\nonumber\\&& \left[{\bf X}_5,{\bf X}_1 \right] =
(2-\frac{\kappa}{2}) {\bf X}_5, \quad\quad \left[{\bf X}_2,{\bf
X}_4 \right] = {\bf X}_5.  \label{commut2}
\end{eqnarray}

In what follows, we will use the above Noether symmetries to
derive exact solutions for this non-minimal model. Before going
further, it should be mentioned that since the equation of motion
\eqref{difeq1}, describing the evolution of the metric potential
$B$, does not depend on the other coordinates, it can be
explicitly solved in terms of $B$. Thus, this metric function will
be obtained as follows
\begin{eqnarray}
B(r) = b_1(r+ b_2)^2. \label{Bsol}
\end{eqnarray}
where $b_1$ and $b_2$ are the integration constants. Without loss
of generality, we choose $B(r)=r^2$ and find that the Noether
symmetry generators (\ref{generator})
 result in the following constants of motion
\begin{eqnarray}\label{DE}
& & I_1 = \,\frac{2fr}{\kappa^2}-\,\frac{r^2f^{'}}{\kappa^2}\,+\,I_{3}\,\phi\,-\,\frac{2}{\kappa^2}\,r\nonumber\nonumber\\
&& I_{2}\,=\,0, \quad\quad\quad\quad \quad  I_{3}\,=\,2\,r^2\,Y(\mathcal{R})\,\phi^{'},
\end{eqnarray}
Considering \eqref{Gauss}, we can set $I_3=-4q$ and then by
solving the system of differential equation, we find that
\begin{eqnarray}\label{non-minimalsol1}
\phi(r)&=& \frac{4q}{r} +\,C_{1}\,\frac{\beta -1}{3\beta + 1}r^{\frac{3\beta+1}{\beta-1}}, \nonumber\\
    f(r) &=& 1+\frac{I_{1}\,\kappa^2}{3\,r}+\frac{ 4\kappa^2q^2} {r^2}  -
    \frac{C_{1}\,(\beta-1)^2}{4\,a_{0}\,\beta(\,3\beta+1)}r^{\frac{2\beta+2}{\beta-1}},
    \hskip 1 cm for \hskip 0.5 cm \beta \neq 0,1,-\frac{1}{3},
\end{eqnarray}
where $C_{1}=\left( {16\beta\,{q}^{2}{k}^{2}a_{0}}
\right)^{\frac{1}{ 1-\beta}} $ is the integration constants. Going
back to minimally coupled Einstein-Maxwell field equations and
checking the solution, we find that the constant of motion $I_{1}$
has a dimension of {\it{mass}}. It is important to mention that
the metric function \eqref{non-minimalsol1} that we get from the
Noether symmetry approach is completely in agreement with the one
that obtained using the method of Lagrange multipliers in
\cite{Dereli:2011gh}. Regarding the excluding values of $\beta$,
we find that by considering the function $Y(\mathcal{R})$ in the
form of \eqref{Y} for $\beta=1$ and $-\frac{1}{3}$ and solving the
system of differential equations \eqref{DE} the following
solutions will be obtained
\begin{itemize}
    \item{$\beta=1$:
    \begin{eqnarray}
    \phi(r)&=& \frac{4q}{r}(1+12C_2a_0), \nonumber\\
    f(r) &=& 1+\frac{I_{1}\,\kappa^2}{3\,r}+\frac{ 4\kappa^2q^2} {r^2}
    (1+12C_2a_0)+C_2r^2,
    \end{eqnarray}}
where $C_2$ is an integration constant which can be fixed as $C_2=\frac{a_0}{q^4}$.
        \item{$\beta=-\frac{1}{3}$:
\begin{eqnarray}
            \phi(r)&=& -\frac{4q}{r},  \nonumber\\
            f(r) &=& 1+\frac{I_{1}\,\kappa^2}{3\,r}-\frac{ 4\kappa^2q^2} {r^2}
            -\frac{r^2}{96a_0},
\end{eqnarray}}
\end{itemize}
An important point that matters concerned with the obtained
solution is the fact that we expect to get the
Reissner-Nordstr\"{o}m (RN) solution in the asymptotic behavior of
parameter $a_0$. In this way, the obtained solution for $\beta=1$
satisfies the expected behavior. However, the solution regarding
$\beta=-\frac{1}{3}$ does not meet the well-behaved asymptotic RN
behavior and hence we do not consider it as an appropriate
solution for the theory.  In the following, we will study the
obtained solution of the theory for $\beta \neq 0,1,-\frac{1}{3}$.

Before going further, it is worth mentioning one point concerning
the physical importance of the obtained solution. Indeed, choosing
some special values of parameters can result in explaining the
effects of dark matter. Let us consider the region where $r$ is
large. Hence, for $-1<\beta<-\frac{1}{3}$, the last term of $f(r)$
in \eqref{non-minimalsol1} dominates and we can approximate $f(r)$
as
\begin{equation}\label{dark}
     f(r) \approx 1 - \mathcal{C}
     r^{-\gamma},
     \end{equation}
where $\mathcal{C}=\frac{C_{1}\,(\beta-1)^2}{4\,
a_{0}\,\beta(\,3\beta+1)}$ and $\gamma=
{\frac{2\beta+2}{1-\beta}}$. One can choose some values of
parameters $\beta$ and $a_0$ such that we have $\mathcal{C}<0$ and
$0<\gamma<1$. Therefore, \eqref{dark} shows that the matter
particle experiences the potential proportional to $r^{-\gamma}$,
which means the force strength is proportional to $r^{-1-\gamma}$.
$\mathcal{C}<0$ guarantees the force is attractive and for
$0<\gamma<1$ the force is stronger than the one produced by
Newton's potential for large $r$. Hence, these solutions might
explain the rotation curves of the galaxies and some effects of
dark matter.

\subsection{Properties of the solution}
Here, we investigate different properties of the obtained metric
function \eqref{non-minimalsol1}. Before we start, it should be
noted that the existence of (singular) black holes has two
conditions: the presence of curvature singularity and the
existence of an event horizon covering it. To begin with, we
examine the existence of the singularity for the obtained
solutions. In this regard, we calculate the Kretschmann scalar and
find that
\begin{eqnarray}
\mathcal{K} =&&\mathcal{R}_{abcd}\mathcal{R}^{abcd}=\nonumber\\
&&{\frac {1}{\mathcal{A}}\Big\{
    16^{{\frac {\beta}{\beta-1}}}\Big[ m \left( \beta+3 \right) r-2\,{
    q}^{2} \left( 5\,\beta+7 \right)\Big] \left( 3\,\beta+1
\right)a_{0} \beta r^{{\frac
{8-4\beta}{\beta-1}}}+\mathcal{B}\left(\frac{r^{8}}{ \beta\,a_{0}
    \,{q}^{2}}\right) ^{\frac{1}{1-\beta}}\Big\} }\nonumber\\
&&+{ \frac
{16}{{r}^{8}}}\Big(3\,{m}^{2}{r}^{2}-24\,m{q}^{2}r+56\,{q}^{4}\Big),\label{Kretschmann}
\end{eqnarray}
where $\mathcal{A}=2( 3\,\beta+1) ^{2}{\beta}^{2}{a_{0}}^{2} [
256\,\beta\,a_{0}\,{q}^{2}] ^{1/ ( \beta-1)}$ and
$\mathcal{B}=3\,{\beta}^{4}+2\,
{\beta}^{3}+10\,{\beta}^{2}+10\,\beta+7$. According to this
relation, we find that the Kretschmann scalar diverges at the
origin, and hence, there is a curvature singularity located at
$r=0$. However, the asymptotic behavior of this curvature
invariant at infinity depends on the value of parameter $\beta$ as
follows
\begin{equation}\label{Kinfinity}
\lim\limits_{r\rightarrow \infty}\,\mathcal{K}=\left\{
\begin{array}{cc}
0 & \,\,\,\,\,\,\,\,\,\,\,\,\,\,\beta<1\\\\
\infty & \,\,\,\,\,\,\,\,\,\,\,\,\,\,\beta>1
\end{array}
\right.
\end{equation}

Besides, studying the behavior of the Kretschmann invariant in
large $\beta$ shows that the curvature of spacetime has a nonzero
finite value when $r$ goes to infinity:
\begin{equation}
\lim\limits_{\beta\rightarrow \infty}\,\mathcal{K}=
\frac{1}{6\alpha^2}+\frac{16}{r^8}\,\left(3\,{m}^{2}
{r}^{2}-24\,m{q}^{2}r+56\,{q} ^{4}\right)+O \left(\frac{1}{\beta}
\right).
\end{equation}

Now, we try to study some physical properties of the solution in
special case. First of all, the basic features of the spacetime
geometry is completely dependent on the value of the parameter
$\beta$:
 \begin{figure}[b!]
    \centering
    \subfigure[\,$a_{0}=-1$, $\beta=-0.2$ and $m=0.001$]{\includegraphics[scale=0.3]{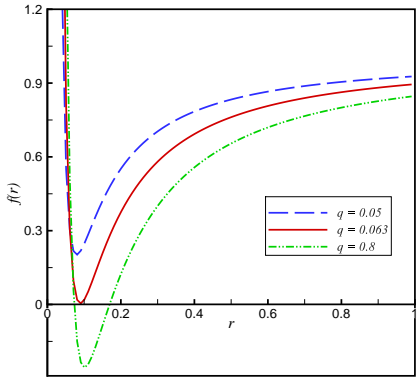}\label{fig1a}} \hspace*{1cm}
    \subfigure[\,$a_{0}=-1$, $\beta=-0.2$ and $q=0.06$
    ]{\includegraphics[scale=0.3]{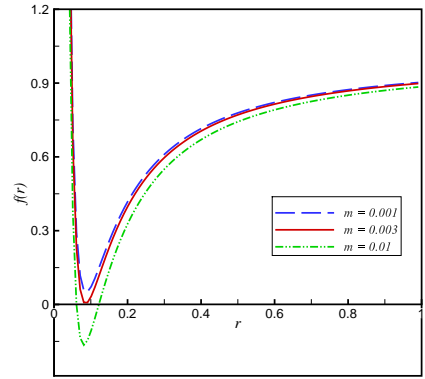}\label{fig1b}} \hspace*{1cm}
    \subfigure[\,$a_{0}=-1$, $q=0.06$ and $m=0.001$
    ]{\includegraphics[scale=0.3]{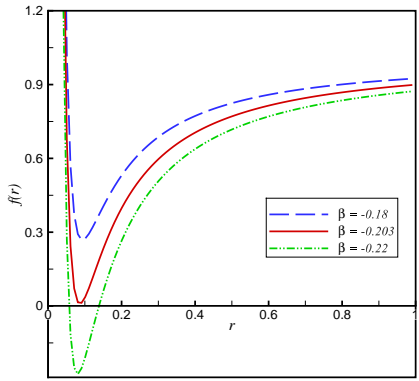}\label{fig1c}}  \hspace*{1cm}
    \subfigure[\, $\beta=-0.2$, $q=0.06$ and $m=0.001$
    ]{\includegraphics[scale=0.3]{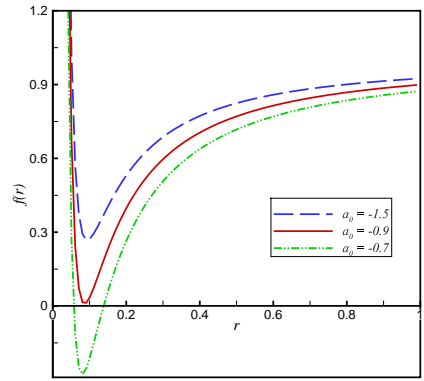}\label{fig1d}}
    \caption{Behavior of $f(r)$ with respect to $r$}
    \label{fig11}
 \end{figure}

\begin{figure}[tbh]
    \centering
    \subfigure[\,$a_{0}=-1$, $\beta=-10$ and $q=1$]{\includegraphics[scale=0.3]{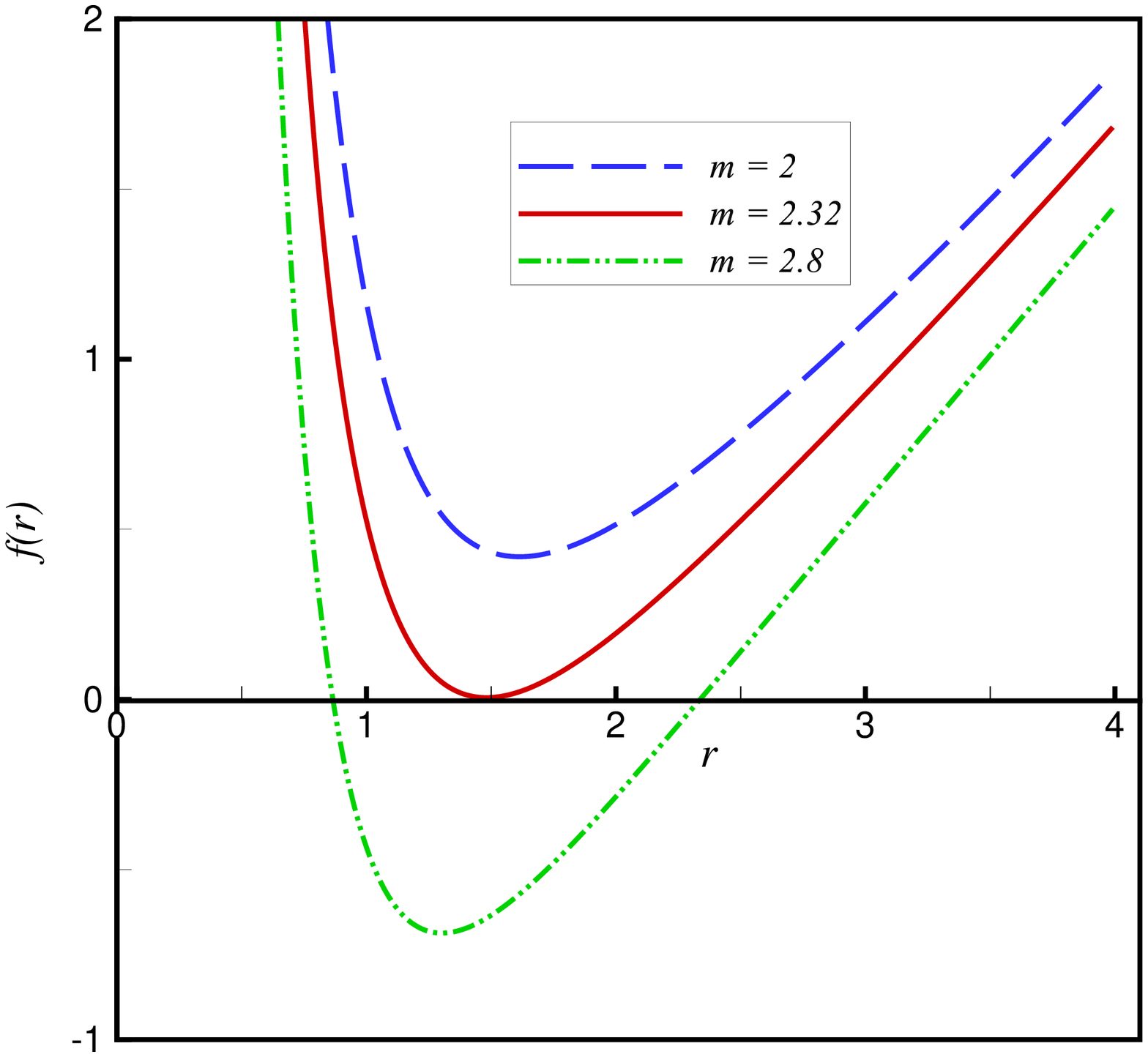}\label{fig1}} \hspace*{1cm}
    \subfigure[\,$a_{0}=-10$, $\beta=-10$ and $m=3$
    ]{\includegraphics[scale=0.3]{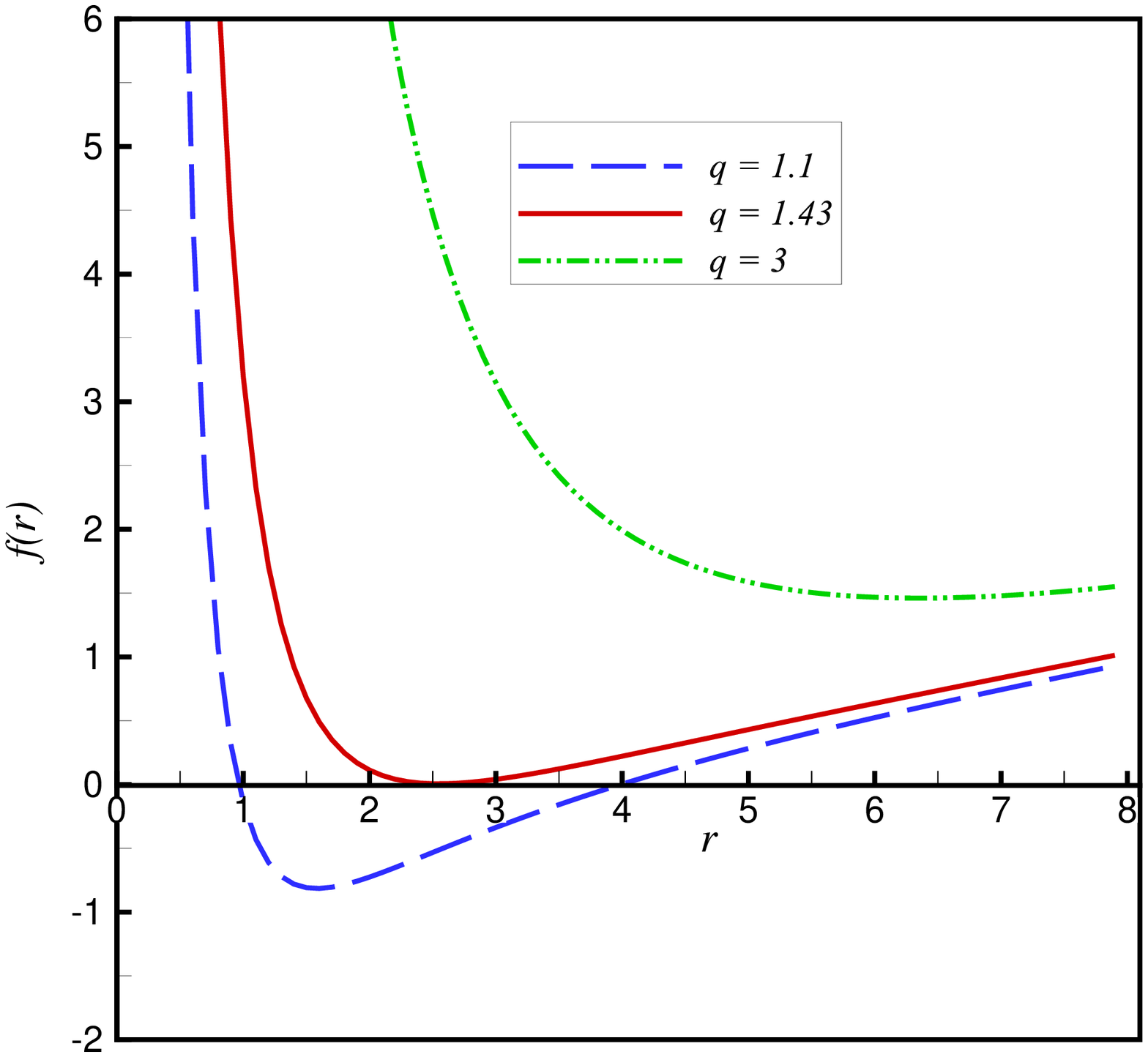}\label{fig2}}
    \caption{Behavior of $f(r)$ with respect to $r$}
    \label{fig22}
\end{figure}

\begin{figure}[!htb]
    \centering  {%
        \includegraphics[scale=0.3
        ]{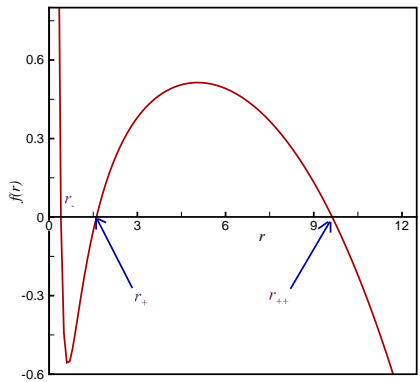}\label{fig3}}
    \caption{Behavior of $f(r)$ with respect to $r$
        for $a_{0}=15$, $\beta=5$ and $m=1$.}\label{fig33}
\end{figure}

\begin{figure}[tbh]
    \centering
    \subfigure[\,$a_{0}=-1$, $\beta=-10$ and $q=1$]{\includegraphics[scale=0.3]{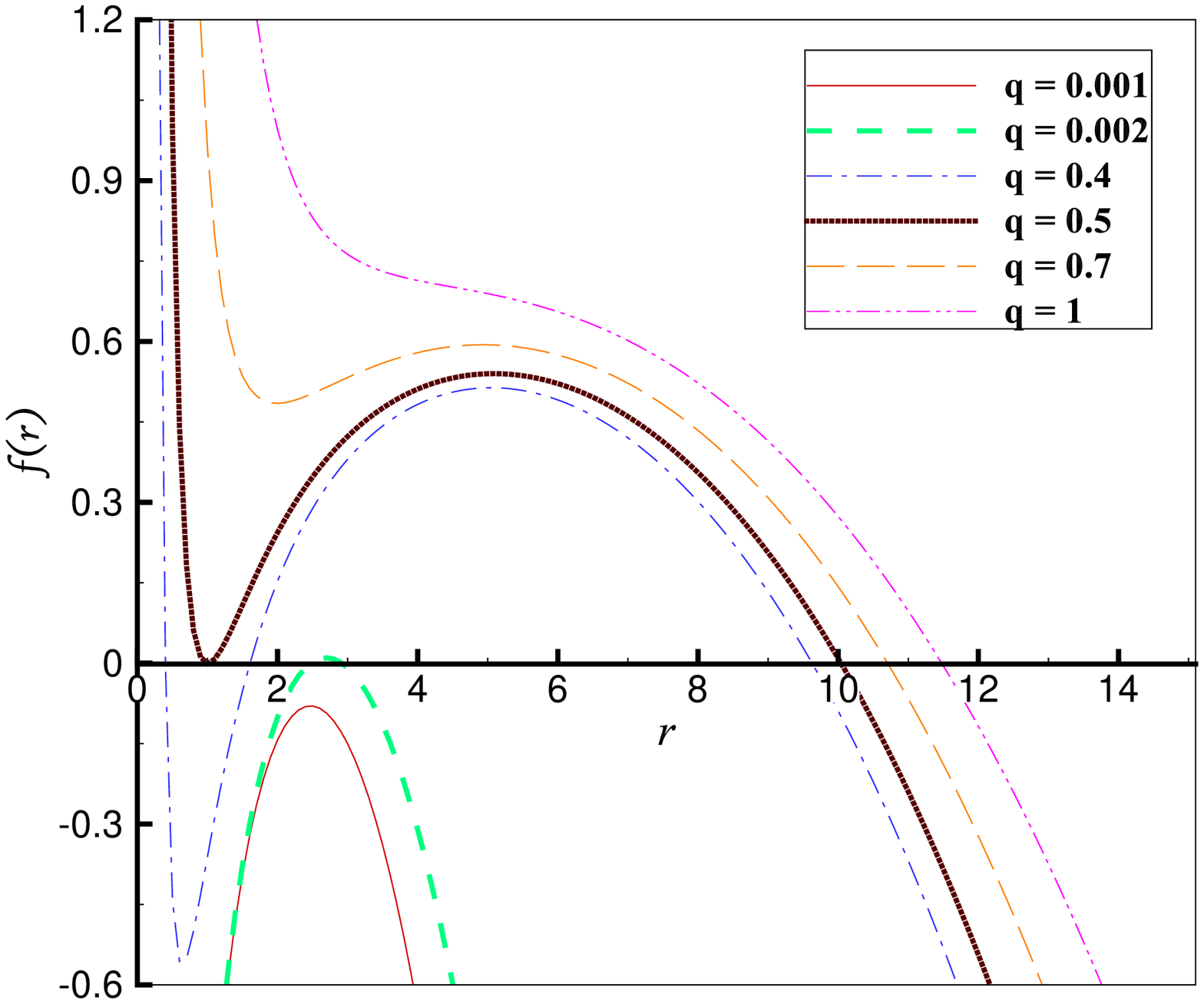}\label{fig4a}} \hspace*{1cm}
    \subfigure[\,$a_{0}=-10$, $\beta=-10$ and $m=3$
    ]{\includegraphics[scale=0.3]{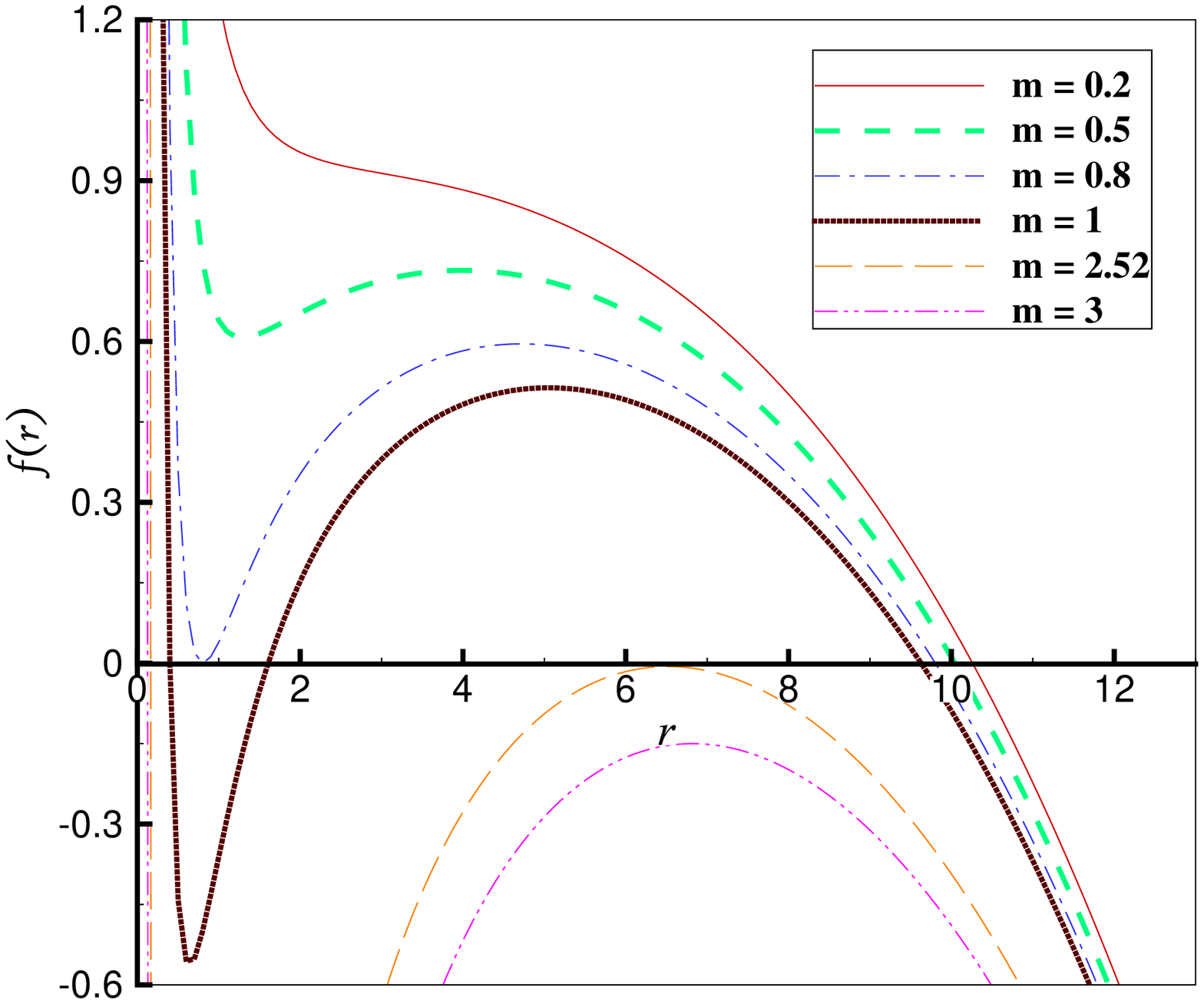}\label{fig4b}}
    \caption{Behavior of $f(r)$ with respect to $r$}
    \label{fig44}
\end{figure}

\begin{figure}[tbh]
    \centering
    \subfigure[\, $\beta=50$, $q=0.5$ and $m=1$]{\includegraphics[scale=0.3]{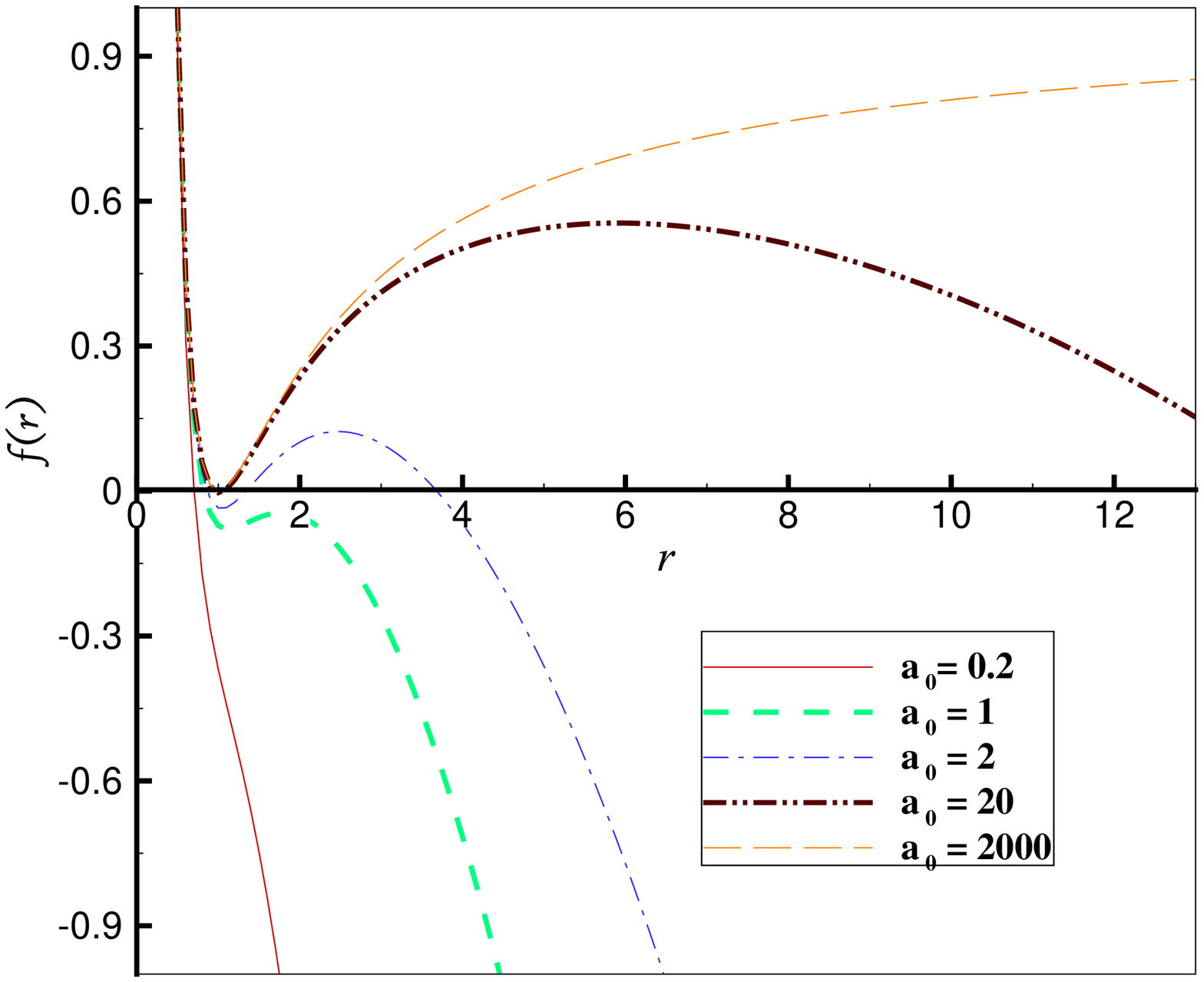}\label{fig4}} \hspace*{1cm}
    \subfigure[\,$a_{0}=1$, $q=0.5$ and $m=1$
    ]{\includegraphics[scale=0.3]{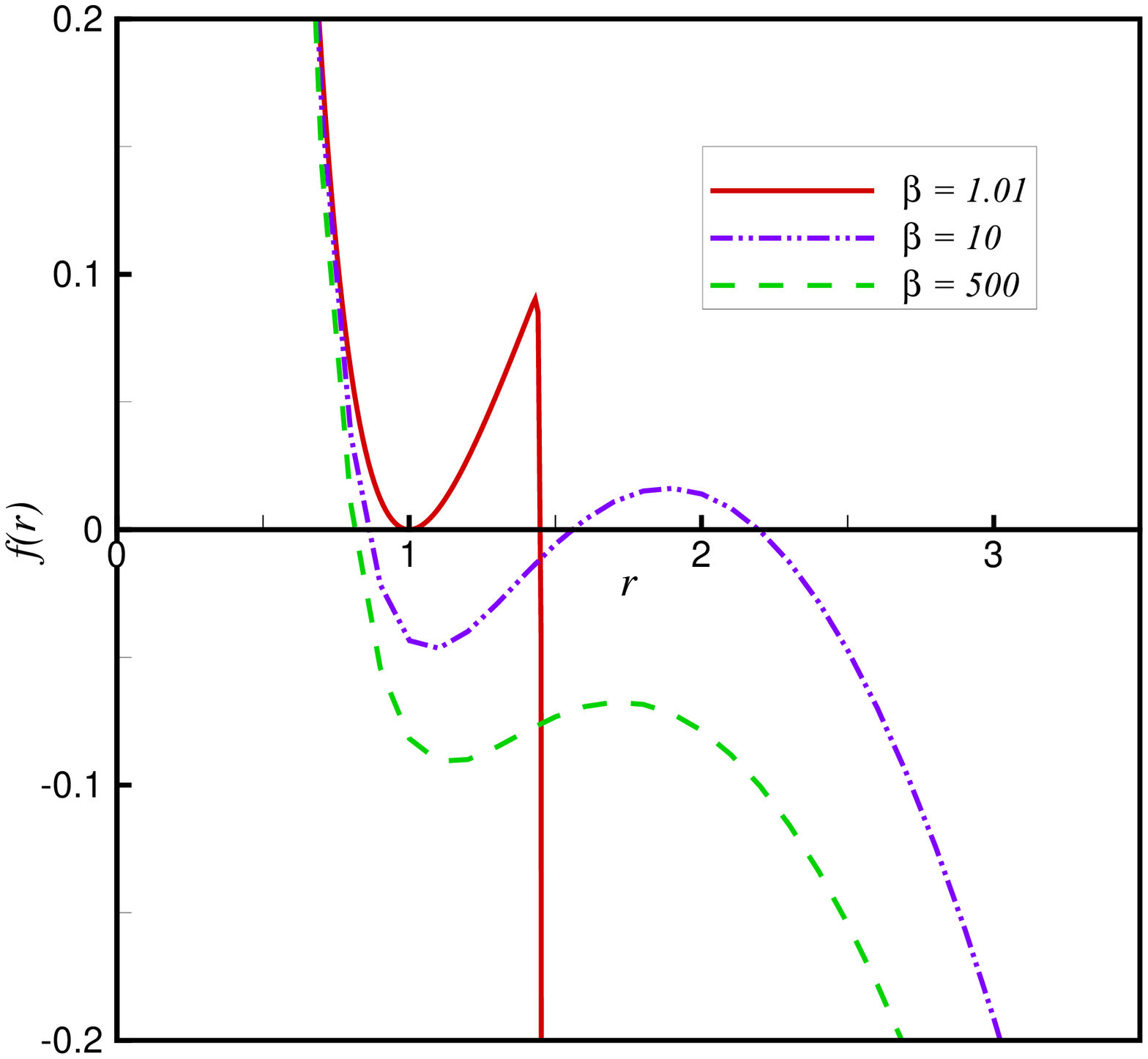}\label{fig5}}
    \caption{Behavior of $f(r)$ with respect to $r$}
    \label{fig55}
\end{figure}

\begin{itemize}

\item{The obtained solution tends to the Schwarzschild solution for $q=0$ and
$\beta<1$.}

\item{This solution is asymptotically flat for the values of $\beta\neq 0, -\frac{1}{3}$ in the range of
$-1<\beta<1$. For this case, the behavior of $f(r)$ is sketched
for different model parameters in Figs \ref{fig11}. According to
these figures one finds that by selecting the value of parameter
$\beta$ from the mentioned interval and depending on the other
metric parameters, this solution could represent a black hole with
two horizons, an extreme black hole with a degenerate horizon or a
naked singularity.}

\item{ Regarding $\beta<-1$, the solution is neither asymptotically flat nor
(A)dS. Strictly speaking, although the Kretschmann scalar vanishes
in large $r$, the metric function is not asymptotically flat due
to the fourth term of Eq. (\ref{non-minimalsol1}) which leads
$f(r)$ to tend to infinity as $r\rightarrow\infty$. The behavior
of the corresponding metric function has been shown in Fig.
\ref{fig22}. Regarding this figure, we find that increasing
(decreasing) the value of parameter $m$, ($q$) leads to changing
the number of horizons from zero to two.}

\item{Choosing the parameter $\beta$ in the range of $\beta>1$
results in black holes with cosmological horizon. Indeed, solution
with this condition meets the de Sitter (dS)-like behavior at
$r\rightarrow\infty$. Also, depending on the other metric
parameters, the corresponding black hole could admit up to three
horizons: a cosmological horizon $r_{++}$, a black hole horizon
$r_{+}$, and an internal horizon $r_{-}$. To be more clear, we
have provided Fig. \ref{fig33} in which the behavior of the metric
function in terms of $r$ has been plotted for a set of model
parameters. However, the number of horizon can change by varying
the values of model parameters which are depicted in Figs.
\ref{fig44}. According to this figures, two horizons $r_{-}$ and
$r_{+}$ coincide at a certain value of parameters. Besides, for
another set of model parameters, the other two horizons $r_{+}$
and $r_{++}$ will coincide each other. We also investigate the
effect of increasing the value of parameters $a_{0}$ and $\beta$
on the geometric feature of the spacetime. To this regard, we have
provided Figs \ref{fig55}. As it is clear from these figures,
increasing (decreasing) the value of $a_{0}$, ($\beta$) leads to
increasing the number of horizon from one to two (two to three).}
 \end{itemize}

\section{Black Hole Thermodynamics and Thermal Stability}\label{4}

\begin{figure}[!htb]
    \centering  {%
        \includegraphics[scale=0.3
        ]{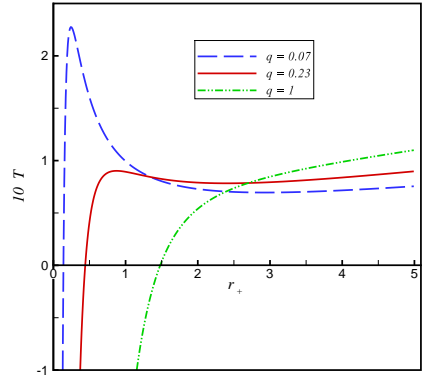}}
    \caption{Behavior of $T$ with respect to $r_{+}$
        for $a_{0}=-1$ and $\beta=-10$.}\label{figT1}
\end{figure}

\begin{figure}[tbh]
    \centering
    \subfigure[\,$q=0.2$]{\includegraphics[scale=0.3]{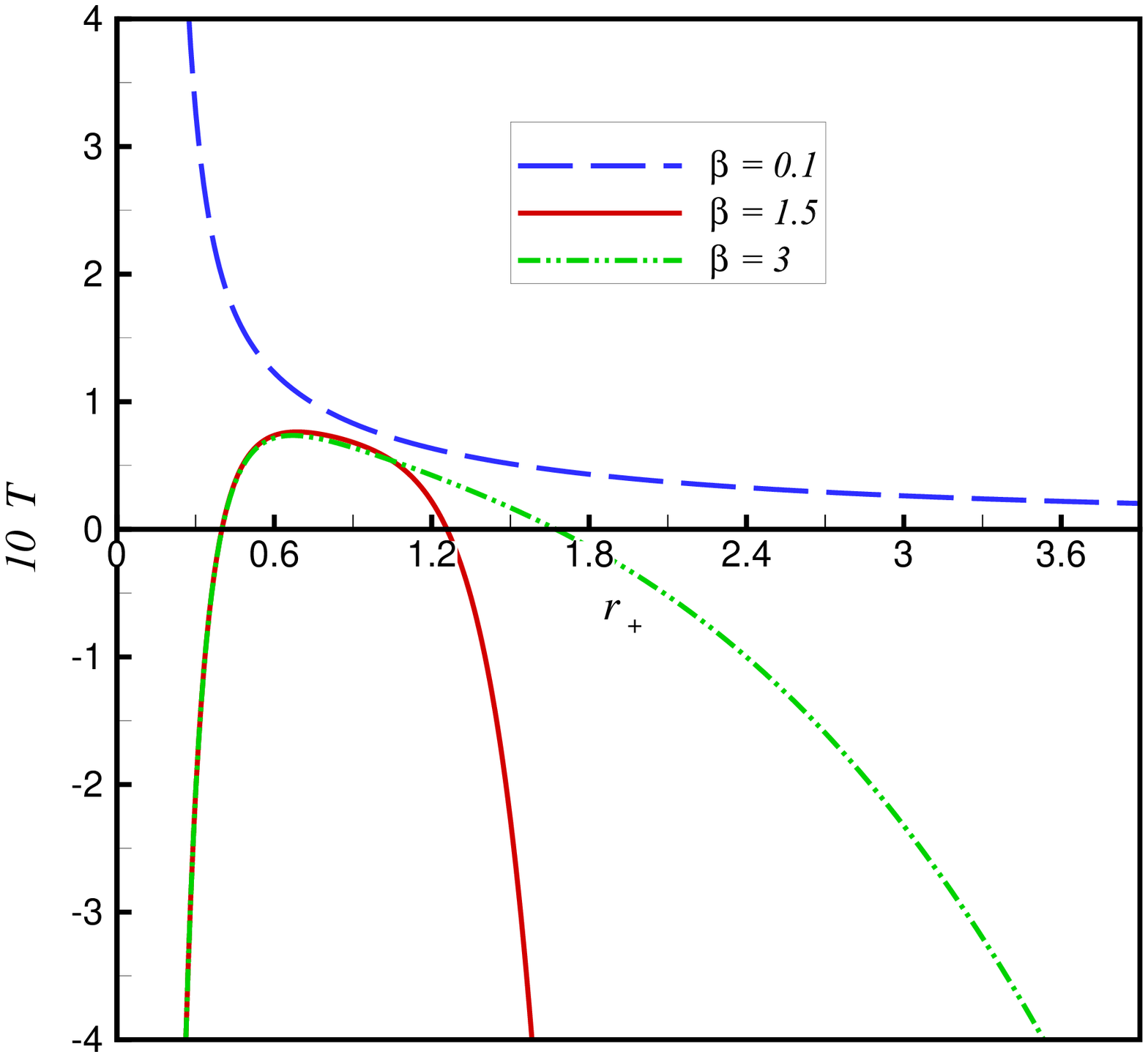}\label{figT3}} \hspace*{1cm}
    \subfigure[\,$q=2$
    ]{\includegraphics[scale=0.3]{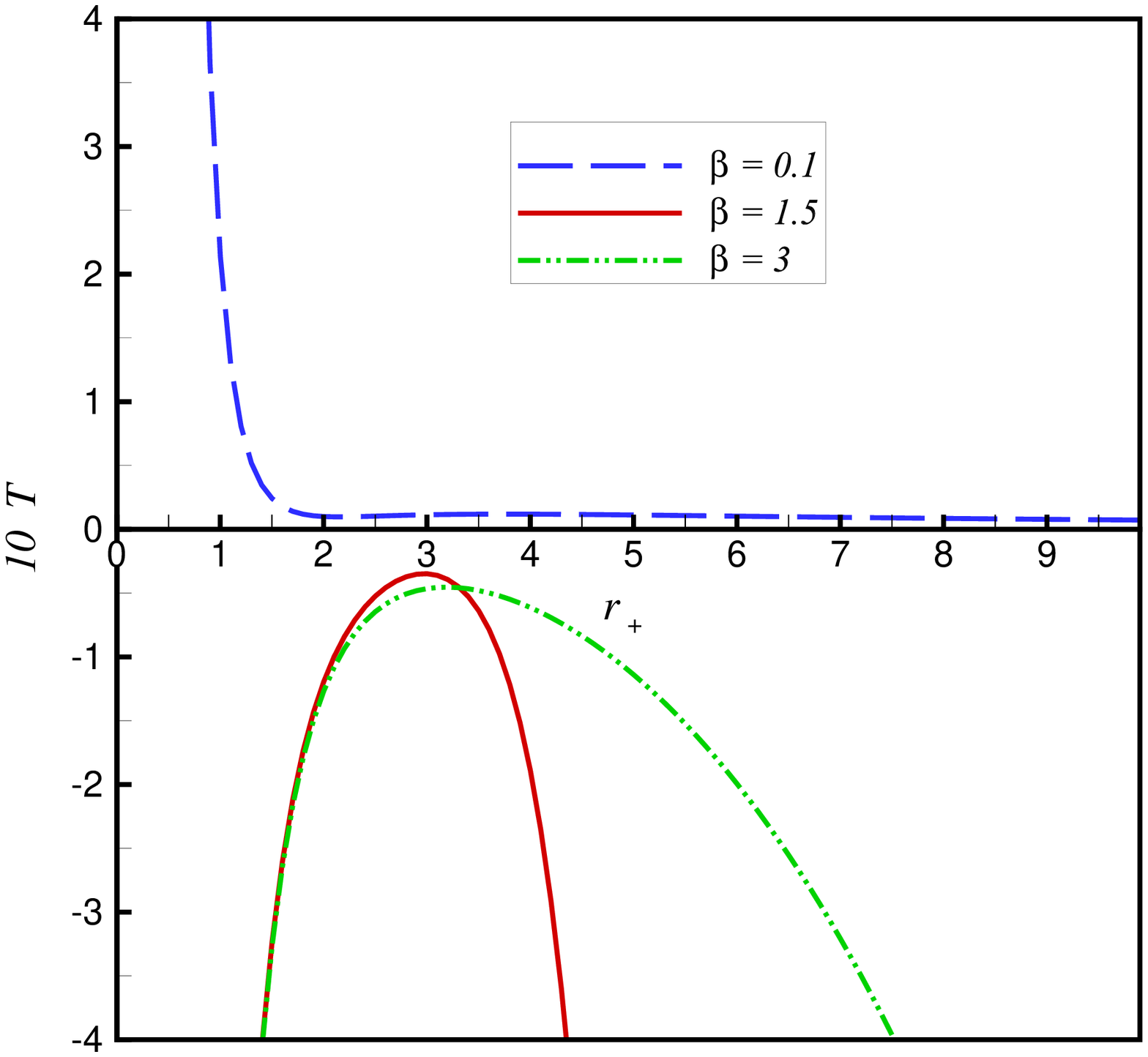}\label{figT4}}
    \caption{Behavior of $T$ with respect to $r_{+}$ for $a_{0}=1$}
    \label{figT}
\end{figure}

One of the most important steps in the transition from the classic
aspect of gravitation to quantum gravity is analyzing black hole
thermodynamics. To study the thermodynamic properties of the
solution, the starting point is the temperature calculation. The
Hawking temperature of the black holes at the event horizon
$r_{+}$ can be obtained through the surface gravity interpretation
\cite{hawking1}
\begin{eqnarray}\label{tem}
T=\frac{1}{4\pi\,r_{+}}-{\frac {{q}^{2}}{\pi r_{+}^{3} }}-\,\frac
{q^2(\beta-1)\; r_{+}^{{\frac {\beta+3}{\beta-1}}} }{\pi\left(
16\,\beta\,a_{0}\,{q}^ {2} \right) ^{ \left( \frac{\beta}{\beta-1}
\right) }}.
\end{eqnarray}
The typical behavior of temperature is depicted in Figs.
\ref{figT1} and \ref{figT} for different parameters of our model.
Calculations show that for the negative values of $a_0$ and
$\beta$ the temperature is not positive for all horizon radii, and
therefore, there will be a constrain on the $r_{+}$ to have the
physical solution as it is shown in Fig. \ref{figT1}. Besides, the
effect of changing the parameter $q$ has been investigated for
this case. As shown in this figure, by increasing $q$, the
starting point of physically acceptable range of black hole will
be at a larger horizon radius. Concerning the positive values of
$a_0$ and $\beta$, we find that depending on the parameters
$\beta$ and $q$, the positive range of temperature can change.
According to Fig. \ref{figT}, it is shown that we encounter to
three different cases: the Hawking temperature can be positive
everywhere, it can be limited to a specific range, or we have no
acceptable physical range.

\begin{figure}[h!]
    \centering
    \subfigure[\,$a_{0}=-1$, $\beta=-1$ and $q=0.07$]{\includegraphics[scale=0.3]{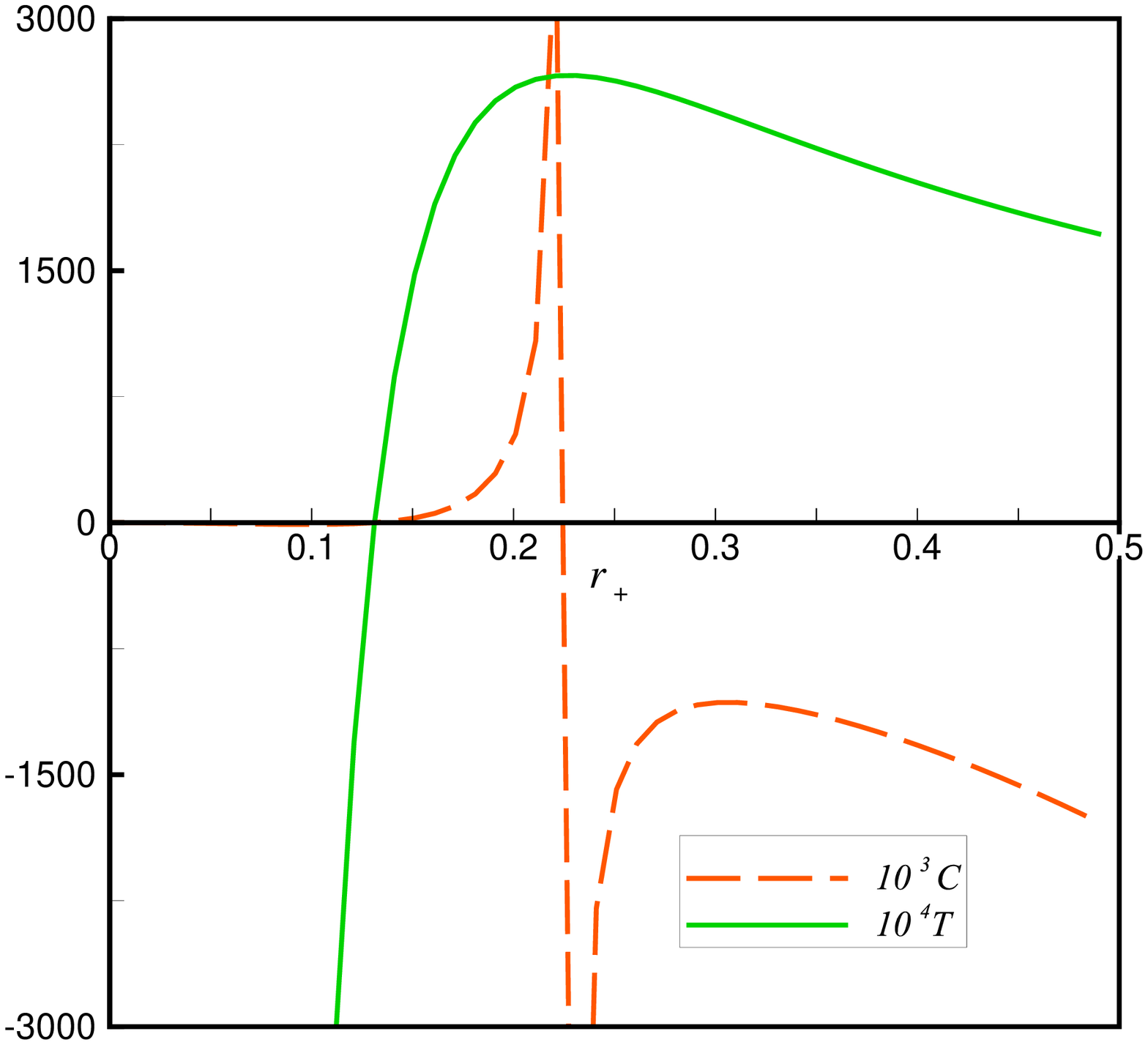}\label{c1}} \hspace*{1.2cm}
    \subfigure[\,$a_{0}=1$, $\beta=1.1$ and $q=0.07$
    ]{\includegraphics[scale=0.32]{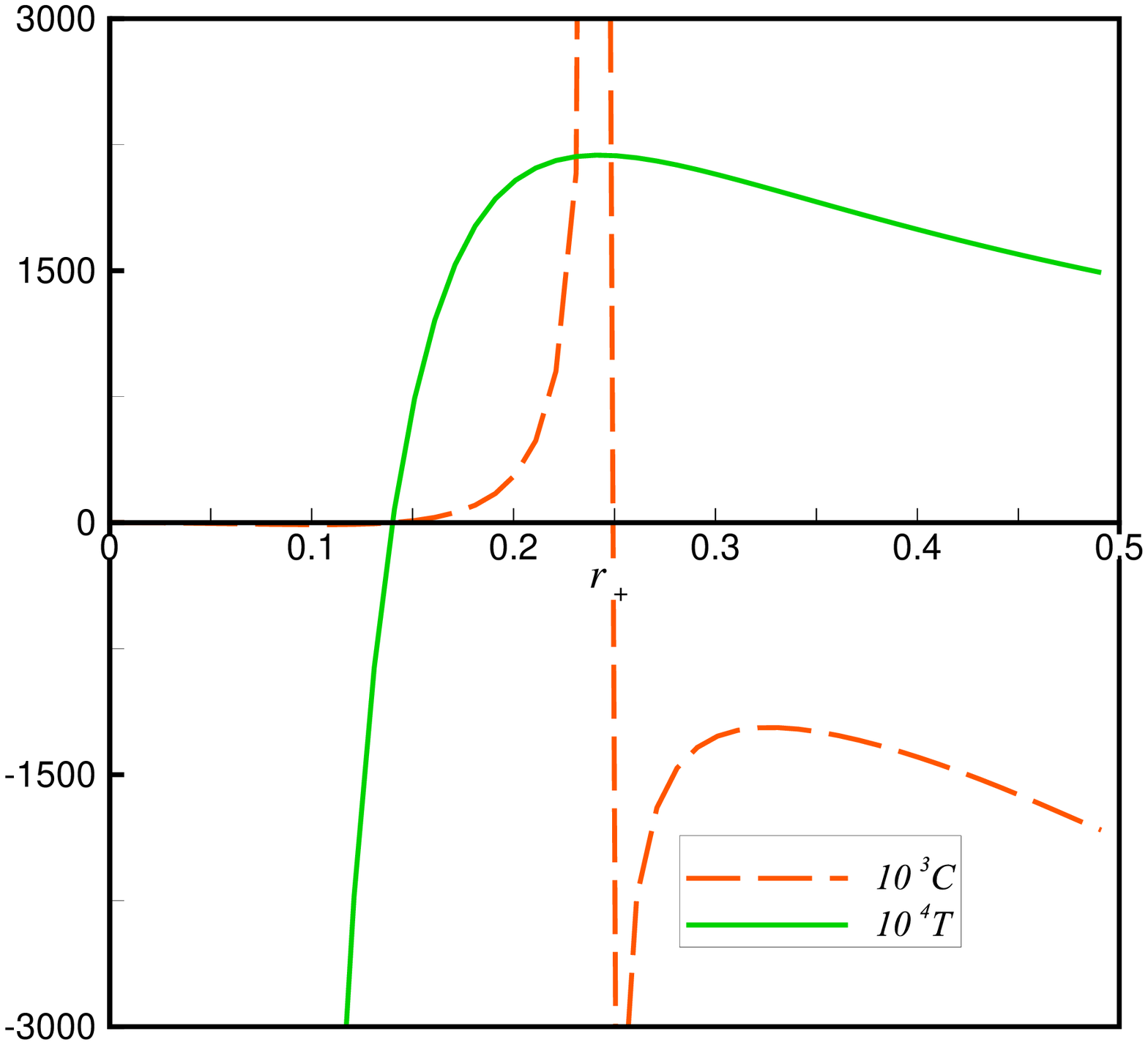}\label{c2}} \hspace*{1cm}
    \subfigure[\,$\beta=10$ and $q=1$]{\includegraphics[scale=0.32]{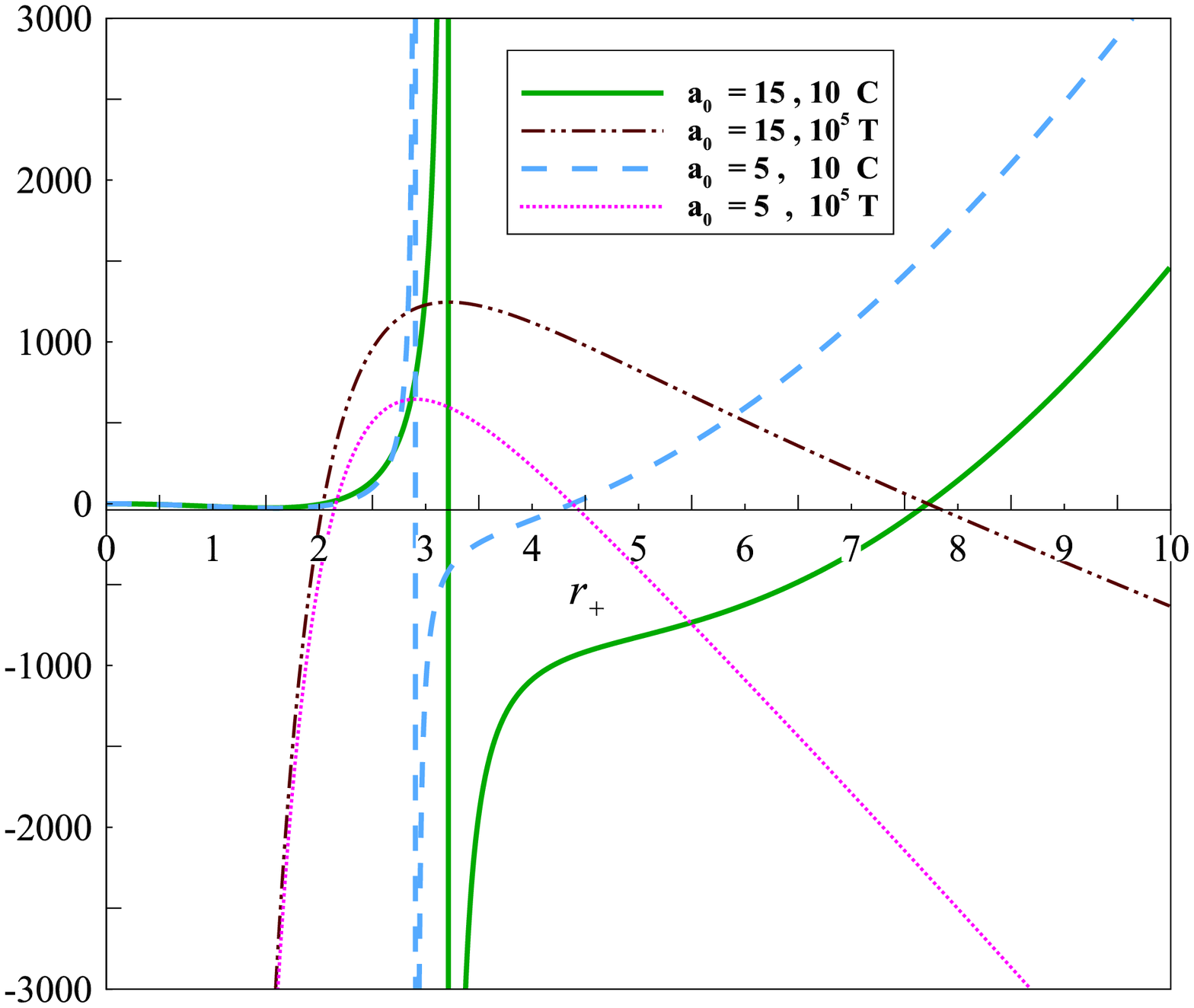}\label{c3}}
    \hspace*{1cm}
    \subfigure[\,$a_{0}=-0.2$, $\beta=-10$ and $q=0.09$]{\includegraphics[scale=0.32]{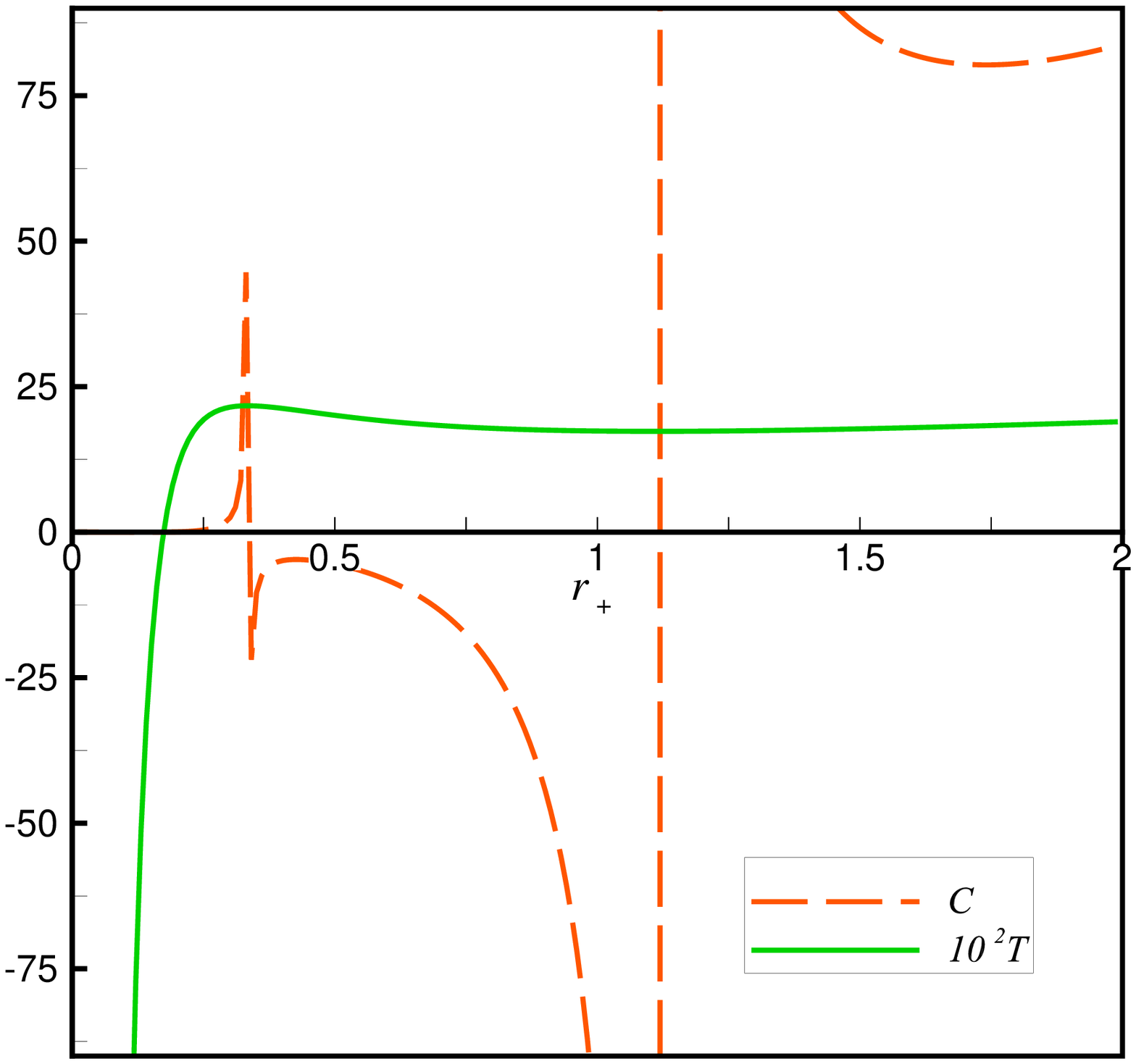}\label{cnew}}
    \caption{Behavior of $C$ with respect to $r_{+}$}
    \label{c}
\end{figure}

The next interesting quantity in black hole thermodynamic context
is entropy that can be obtained by using the area law in the case
of Einsteinian black holes \cite{entropy1, entropy2}. Therefore,
the entropy of the obtained black hole becomes
\begin{equation}
S=\pi r_{+}^2.
\end{equation}

Here, it is necessary to pause and explain about using the area
law relation for the entropy; Indeed, the direct coupling between
matter and geometry components results in the fact that the matter
stress tensor is no longer conserved and there is an energy
transfer between the two components. To be more clear, one can get
the field equation for the action \eqref{action}. Variation with
respect to the metric $g_{\mu\nu}$ yields the modified Einstein
equations of motion as follows
\begin{eqnarray}
    \left( 1 - Y_{\mathcal{R}}(\mathcal{R}) {F^2} \right) \left(\mathcal{R}_{\mu\nu} - {1 \over 2} g_{\mu\nu} \mathcal{R} \right) &=& -8\pi G Y_{\mathcal{R}}(\mathcal{R}) T_{\mu\nu} + {1 \over 2} \left[\mathcal{R} - \left(1 - Y_{\mathcal{R}}(\mathcal{R} \right)F^2 )\right] g_{\mu\nu} \nonumber\\
    &&+  \left(  \nabla_\mu \nabla_\nu - g_{\mu\nu} \square \right) (1 - Y_{\mathcal{R}}(\mathcal{R})F^2), \label{Field}
\end{eqnarray}
in which the matter energy-momentum tensor is, as usually, defined by
\begin{equation}
     T_{\mu\nu} = -{2 \over \sqrt{-g}} {\left(\sqrt{-g} F^2\right) \over  g^{\mu\nu}}.
\end{equation}
Now, taking into account the covariant derivative of the obtained
field equation and the Bianchi identities as well as the identity
$\left( \square \nabla_\nu - \nabla_\nu \square
\right)Y_{\mathcal{R}}(\mathcal{R})= R_{\mu\nu} \nabla^\mu
Y_{\mathcal{R}}(\mathcal{R})$, imply the non-(covariant)
conservation law
\begin{equation}
    \label{cons} \nabla^\mu T_{\mu\nu} = { Y_{\mathcal{R}}(\mathcal{R}) \over Y(\mathcal{R})} \left( g_{\mu\nu} F^2 - T_{\mu\nu} \right) \nabla^\mu \mathcal{R} ~~,
\end{equation}
and, as expected, in the  limit $Y_{\mathcal{R}}(\mathcal{R}) =
1$, one recovers the conservation law $\nabla^\mu T_{\mu\nu} = 0$.
Due to the energy transfer between the two components, the system
will be a dynamical system. It has been shown that one way to
investigate the thermodynamic properties of dynamical black hole
is working in the limit of weak gravity and consider the
non-minimal term as a perturbation around the Einstein gravity,
known as the linearized amplitude expansion \cite{Biswas:2022grc}.
In the context of the field equation, , we can recast the field
equation \eqref{Field} in a way that the higher order corrections
are written as an energy-momentum tensor of geometrical origin
describing an effective source term on the right hand side of the
standard Einstein field equations, namely
\begin{equation}
    G_{\mu\nu}=T_{\mu\nu}^{m}+T_{\mu\nu}^{g},
\end{equation}
where
\begin{equation}
    T_{\mu\nu}^{m}=\frac{-8\pi G Y_{\mathcal{R}}(\mathcal{R})}{1 - Y_{\mathcal{R}}(\mathcal{R}) {F^2} }T_{\mu\nu},
\end{equation}
and
\begin{equation}
T_{\mu\nu}^{g}=\frac{1}{1 - Y_{\mathcal{R}}(\mathcal{R}) {F^2} }\left[{1 \over 2} \left[\mathcal{R} - \left(1 - Y_{\mathcal{R}}(\mathcal{R} \right)F^2 )\right] g_{\mu\nu} + \left(  \nabla_\mu \nabla_\nu - g_{\mu\nu} \square \right) (1 - Y_{\mathcal{R}}(\mathcal{R})F^2)\right].
\end{equation}
Therefore, it seems reasonable to use area law formula for the
entropy of the black hole in this model and examine it via the
first law and Smarr formula.

In order to calculate the electric potential of the black hole, we
use the fact that it is the electrostatic potential difference
between the horizon and the boundary at infinity \cite{haj}
\begin{equation}
\Phi=A_\mu\chi^\mu\Big|_{r\to\infty}-A_\mu\chi^\mu\Big|_{r= r_{_+}}.
\end{equation}
Hence, by considering the Killing vector $\chi^\mu=(-1,0,0,0)$ as
the null generator of the horizon, the electric potential of the
black hole is obtained as follows
\begin{equation}
\Phi(r_{_+})=\frac{4q}{r_{_+}} +\,C_{1}\,\frac{\beta -1}{3\beta +
1}r_{_+}^{\frac{3\beta+1}{\beta-1}}. \nonumber \label{phi}
\end{equation}
Moreover, for obtaining the mass of solution, one can evaluate the
metric function on its event horizon $f(r=r_{_+})=0$ which results
in the following relation for our black hole solution
\begin{equation}
M(r_{_+},q)=\frac{r_{_+}}{2}+{
    \frac {2{q}^{2}}{r_{_+}}}-{\frac { \left( \beta-1 \right) ^{2} \left( 16\,\beta\,a_{0}\,{
            q}^{2} \right) ^{ \frac{1}{\left(1 -\beta\right)} }}{8a_{0}\,\beta\,    \left( 3\,\beta+1 \right) }{r_{_+}}^{{\frac {3\,\beta+1}{\beta-1}}}}.
\end{equation}

Now, it is logical to examine the validity of the first law of
black hole thermodynamics and the Smarr formula for the obtained
black hole. It is a matter of calculation to show the Smarr
formula is
\begin{equation}\label{smarr}
M=2TS+\Phi Q+2\mathcal{V}a_0,
\end{equation}
where the electric charge per unit volume $Q$ can be found by
calculating the flux of the electric field at infinity which
yields $Q={q}/{4\pi}$. Besides, this relation states that in
addition to $S$ and $Q$, $a_{0}$ is also a thermodynamic variable
which its conjugate will be as follows
\begin{equation}
\mathcal{V}={\frac {\beta-1}{ 8{a_0}^{2}\left( 3\,\beta+1 \right) }
    \left(\frac{1}{16 \beta\,a_0\,{q}^{2}} \right) ^{ \frac{1}{ \beta-1 }}}r_{+}^{{
        \frac {3\,\beta+1}{\beta-1}}},   \label{V}
\end{equation}
Here, it should be mentioned that the third term in the Smarr
formula originates in the non-minimal term in the action. Having
conserved and thermodynamic quantities, it s easy to show that the
obtained thermodynamic quantities satisfy the first law of
thermodynamics
\begin{equation}
dM=T dS+\Phi dQ+\mathcal{V}da_0.
\end{equation}

As a next step, we are going to investigate thermal stability and
look for possible phase transition of our solution. In the
canonical ensemble, the positivity of heat capacity ensures the
local stability in regions where black hole temperature is
positive, while the opposite stands for the unstable black holes
which may undergo phase transition to be stabilized. Regarding the
obtained black hole solution, the heat capacity function is
calculated as
\begin{equation}
C=T\left( \frac{\partial T}{\partial S}\right)^{-1}=\frac{2 \pi
r_{+}^{2} \left[ 4\beta a_{0}\, \left( 4 {q}^{2}-r_{+}^{2}
    \right) + \left( 16 \beta a_{0}\,{q}^{2} \right) ^{ \frac{1}{ 1-\beta}} \left( \beta-1 \right) {r_{+}}^{{\frac {4\beta}{\beta-1
    }}} \right]  }{ -4 \beta a_{0}\, \left( 12\,{q}^{2}-r_{+}^{2}
    \right) + \left( 16 \beta a_{0}\,{q}^{2} \right)^{ \frac{1}{ 1-\beta }} \left( \beta+3 \right) {r_{+}}^{{\frac {4\beta}{\beta-1
    }}}}.
\end{equation}

It is well known that in the presence of positive temperature, the
points at which black hole heat capacity diverges (vanishes) may
correspond to the phase transition. Since the temperature is a
smooth polynomial function and does not blow up, the divergence
point of heat capacity is located at the points of vanishing its
denominator. For arbitrary values of $\beta$ it is not possible to
find, analytically, the root(s) of heat capacity and its
denominator, and therefore, we have to provide numerical analysis
as is shown in Figs. \ref{c} for more details.

Regarding the positive numbers of parameters $a_{0}$ and $\beta$,
the heat capacity experiences one divergence point in the presence
of positive $T$ (see Fig. \ref{c}). We call this special point as
$ {r_{+}}_d$. Therefore, if the radius of the event horizon is
between root of temperature and the divergence point of the heat
capacity, the black hole is thermally stable and meets a necessary
criterion for viable solutions. On the other hand, black holes
whose horizon radius are larger than ${r_{+}}_{d}$ experience the
negative heat capacity which means that a phase transition occurs
at $r_{+}={r_{+}}_d$ between physical and unphysical black holes.
The mentioned admissible domain of the black hole can be altered
by changing the model parameter $a_{0}$. According to Fig.
\ref{c3}, increasing the values of this parameter leads to make
the admissible domain larger. However, in the case of negative
numbers of parameters $a_0$ and $\beta$, the heat capacity meets
two divergences. It shows that black hole with these parameters
experiences a second order phase transition \cite{davis} between
small and large ones. Therefore, middle black holes are not
allowed for this class of parameters.

Now, by calculating the free energy of the system, we complete our
discussion on criticality. The free energy can be obtained
using the Legendre transformation or calculating the on-shell
action as follows
\begin{equation}
A=M-TS=\frac{r_{+}}{4}+{\frac {3\,{q}^{2}}{r_{+}}}+\frac { \left(
\beta+3 \right)  \left( \beta-1 \right)  \left(
        16\,\beta\,a_{0}\,{q}^{2} \right) ^{\frac{1}{1-\beta}}}{
    16  a_{0}\,\beta\, \left( 3\beta+1 \right) }{r_{+}}^{{\frac {3\,\beta+1}{
                \beta-1}}}.
\end{equation}

\begin{figure}[tbh]
    \centering
    \subfigure[\,$\beta=-1$, $q=0.07$]{\includegraphics[scale=0.3]{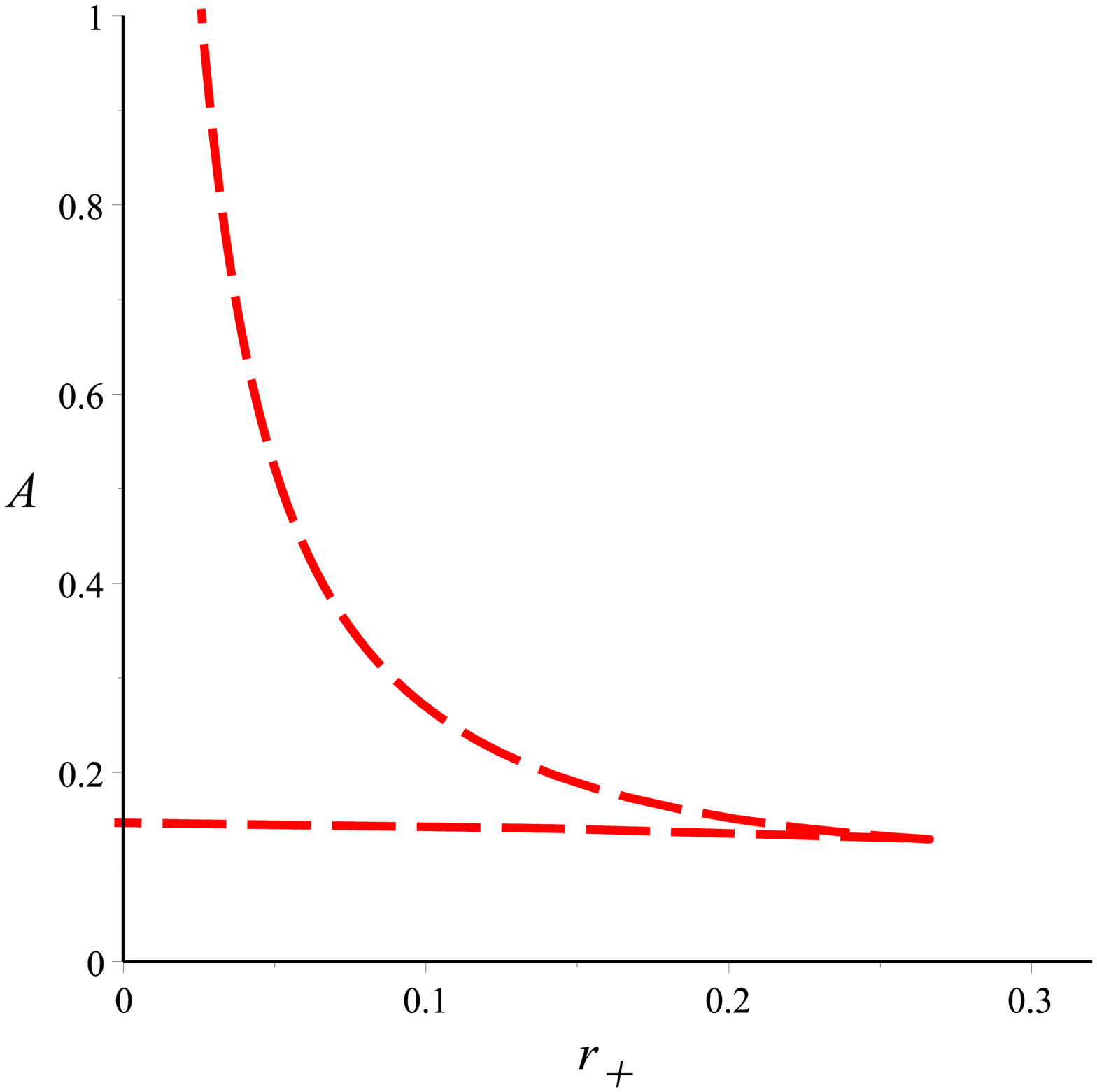}\label{Gne}} \hspace*{1cm}
    \subfigure[\,$\beta=-10$
    ]{\includegraphics[scale=0.3]{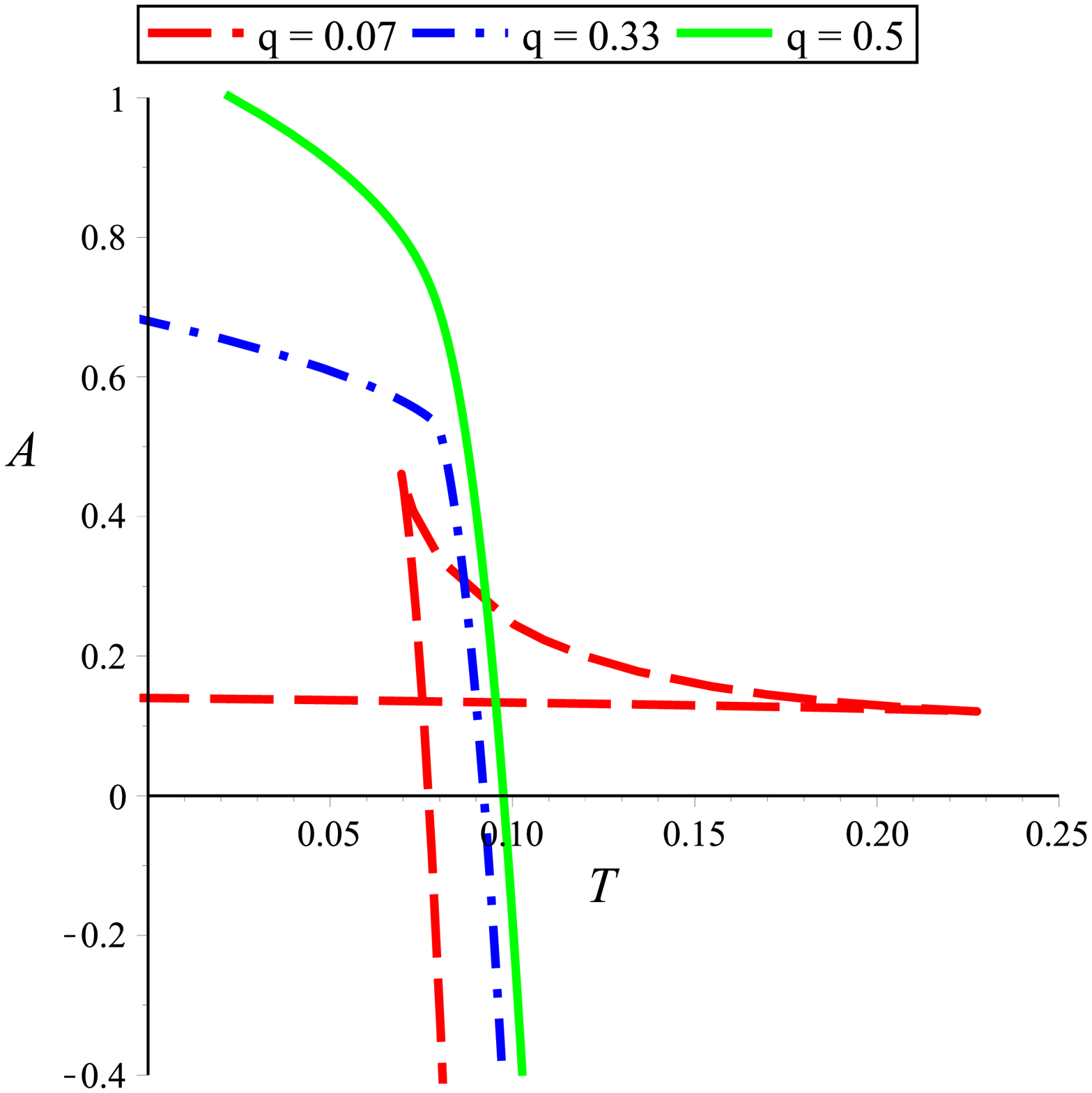}\label{G1}}
    \caption{Behavior of the free energy with respect to temperature
        for $a_{0}=-1$ and $\beta=-10$}
    \label{figG}
\end{figure}

The qualitative behavior of free energy as a function of
temperature is depicted in the Fig. \ref{figG}.


According to Fig. (\ref{figG}), one can see different behavior of
free energy depending on the metric parameters. Fig. (\ref{Gne})
is plotted according to metric parameters of Fig. (\ref{c1}) which
shows that free energy is multi-valued when horizon radius belongs
the interval ($0.2<r_+<0.3$). It is notable that in the mentioned
interval the heat capacity diverges and it changes to negative
value. On the other hand, Fig. (\ref{G1}) shows that the energy is
single-valued for large values of charge parameter and decreases
monotonically with the increasing temperature and is locally
stable. In the case of small $q$, it becomes multi-valued which is
famous as swallow-tail behavior that indicating a first order
phase transition between small and large black holes. It is more
or less the same as van der Waals fluid which in our system occurs
for smaller values of charge parameter. The blue curve indicates a
second-order phase transition where $A$ is single-valued and
continuous, but is non-analytic (differentiable everywhere).

\section{Conclusion}

In this paper, we have studied the black hole solution of a
particular model of non-minimally coupled gravity to
electromagnetic as $Y(\mathcal{R})F^2$. For an arbitrary
$Y(\mathcal{R})$, we have used the Noether symmetry approach to
find five generators and related symmetries with corresponding Lie
algebra. Then, we have chosen a special class of $Y(\mathcal{R})$,
and obtained different generators and related Lie algebra.

Regarding the Noether symmetry approach, we have found exact
solutions of the non-minimal gauge-gravity coupling with black
hole interpretation. The paper was concentrated on two aspects of
these black holes including solutions and geometrical properties
as well as thermodynamic behavior and possible phase transition.

We have considered $\mathcal{R}^\beta F^2$ type non-minimal
coupling between gravity and electromagnetic field. By using the
Noether gauge symmetry approach, we have investigated the possible
symmetries as well as corresponding conserved quantities to find
the exact solutions for this model. The exact metric function was
obtained and it was shown that: I) The Kretschmann scalar admitted
the existence of singularity. II) Depending on the values of the
model parameters $\beta$, $a_{0}$, $q$ and $m$, the obtained
solution could enjoy from zero up to three roots for the metric
function.

The investigations concerning the thermodynamics of these black
holes confirmed: I) Dependency of the temperature on the  model
parameters $\beta$, $a_{0}$, $q$ and $m$. II) The validity of
first law of black hole thermodynamics. III) Existence of
divergency for the heat capacity in the presence of positive
temperature values that indicates the obtained black hole
undergoes second order phase transition. IV) Existence of van der
Waals-like phase transition for the black hole under study.


\section*{Acknowledgments}

S. Mahmoudi is grateful to the Iran Science Elites Federation for the
financial support.

\appendix \section{A Review on Noether Gauge symmetry approach}\label{appendix}

Noether symmetry which is defined in the context of dynamical
systems provides a fascinating procedure to find symmetries of
differential equations and use them to derive an exact analytic
solution of proposed models at a fundamental level, including
modified gravity theories. This approach opens the way to find the
solution either by reducing the dynamic system's degrees of
freedom or determining the unknown functions of the system. The
aim of this appendix is to provide a brief overview on this
approach.

To begin with, we consider the one parameter point transformation,
also called coordinate transformation,
\begin{equation}
    (x,y) \to (\bar{x},\bar{y})
\end{equation}
in which
\begin{equation}
\label{eq1}
\bar{x}=\bar{x}(x,y;\varepsilon),  \qquad \bar{y}=\bar{y}(x,y;\varepsilon).
\end{equation}
where $\varepsilon$ is an arbitrary real parameter. The
first-order Taylor expansion of the infinitesimal transformation
\eqref{eq1} around $\varepsilon=0$ yields
\begin{align}
&\bar{x}(x,y;\varepsilon)=x+\varepsilon\frac{\partial\bar{x}}{\partial\varepsilon}\bigg|_{\varepsilon=0}  %
=x+\varepsilon\xi(x,y)   \label{nsa1.3} \\
&\bar{y}(x,y;\varepsilon)=y+\varepsilon\frac{\partial\bar{y}}{\partial\varepsilon}\bigg|_{\varepsilon=0}  %
=y+\varepsilon\eta(x,y)  \label{nsa1.4} \,.
\end{align}
where the functions $\xi(x,y),\eta(x,y)$, which is called the
\textit{infinitesimal generator} of the transformation, are the
components of the tangent vector $\mathbf{X}$, \textit{i.e.}
\begin{equation}
\label{nsa1.5}
\mathbf{X}=\xi(x,y)\frac{\partial}{\partial x}+\eta(x,y)\frac{\partial}{\partial y}.
\end{equation}
Since our goal is to see how differential equations are affected
by these transformations, we have to extend/prolong them to the
derivatives. Therefore, by means of the following relations
\begin{align}
&\bar{y}' \equiv \frac{d\bar{y}(x,y;\varepsilon)}{d\bar{x}(x,y;\varepsilon)}%
=\frac{y'(\partial\bar{y}/\partial y)+(\partial\bar{y}/\partial x)}{y'(\partial\bar{x}/\partial y)+(\partial\bar{x}/\partial x)}%
=\bar{y}'(x,y,y';\varepsilon),    \label{nsa1.6}  \\
&\bar{y}''\equiv \frac{d\bar{y}'}{d\bar{x}}=\bar{y}''(x,y,y',y'';\varepsilon), \label{nsa1.7} \\
&\ldots \notag
\end{align}
the prolongation/extension of the tangent vector, involving the
$n^{th}$  derivatives, can be computed. Using the first-order
Taylor expanding around $\varepsilon=0$ and substituting Eqs.
\eqref{nsa1.3} and \eqref{nsa1.4} into Eqs. \eqref{nsa1.6} and
\eqref{nsa1.7}, the $n^{th}$ derivatives of the transformed
coordinates are obtained as follows
\begin{align}
&\bar{y}'=y'+\varepsilon\left(\frac{d\eta}{dx}-y'\frac{d\xi}{dx}\right) =y'+\varepsilon\eta^{[1]}  \,,   \\
&\vdots   \notag\\
&\bar{y}^{(n)}=y^{(n)}+\varepsilon\left(\frac{d\eta^{(n-1)}}{dx}-y^{(n)}\frac{d\xi}{dx}\right)  %
=y^{(n)}+\varepsilon\eta^{[n]}  \,,
\end{align}
where
\begin{equation}
\label{nsa1.10}
\eta^{[n]} \equiv \frac{d\eta^{(n-1)}}{dx}-y^{(n)}\frac{d\xi}{dx}=\frac{d^n}{dx^n}(\eta-y'\xi)+y^{(n+1)}\xi \,
\end{equation}
is the $n^{th}$ prolongation function of $\eta$. Therefore, the $n^{th}$ prolongation of the generator $\mathbf{X}$ reads
\begin{equation}
\label{nsa1.12}
\mathbf{X}^{[n]}=\mathbf{X}+\eta^{[1]}\partial_{y'} +...+\eta^{[n]}\partial_{y^{(n)}}.
\end{equation}
It is worth mentioning that we referred only to one parameter
point transformations so far. However, the procedure followed to
define multi parameter point transformations on variables, their
derivatives, as well as their generators, is the same.\par

Now, we are ready to study the behavior of differential equation
under the action of point transformations. A group of point
transformations that maps solutions into another solutions
 is called a \textit{symmetry} of the differential equations.
%
%
%
A specific class of Lie point symmetries are the so-called Noether
symmetries. They are restricted to dynamical systems coming from a
Lagrangian. The Lagrangian function $\mathcal{L}
=\mathcal{L}(t,q^i,\dot{q}^i)$\footnote{The index $i$ takes the
values  $1,2,...,n$ and denotes the number of dimensions of the
configuration space.}, is a function of time $t$, the generalized
coordinates $q^i=q^i(t)$ and their time derivatives $\dot{q}^i(t)$
which contains information about the dynamics of a system. Having
the Lagrangian function in hand, the system's equations of motion
are given by the Euler-Lagrange equations
\begin{equation}
\frac{d}{dt}\left(\frac{\partial \mathcal{L}}{\partial \dot{q}^i}\right) - \frac{\partial \mathcal{L}}{\partial q^i} = 0 \,.
\end{equation}
where it is obtained by the variation of the action integral
\begin{equation}
    A(q^i, \dot{q}^i)=\int_{t_1}^{t_2} \mathcal{L}(t,q^i,\dot{q}^i) dt.
\end{equation}
Noether symmetries use the fact that when we add a total
derivative to the Lagrangian
\begin{equation}
    A'(q'^{i}, \dot{q'}^{i})=A(q^i, \dot{q}^i)+\int_{t_1}^{t_2} \frac{dh(t, q^i, {\dot{q}}^i)}{dt} dt
\end{equation}
the equations of motion do not change and therefore, the set of
solutions remains the same. The function $h$ is called the Noether
or the gauge function.

The Lagrangian-related Noether vector $X$ and the first
prolongation vector field $X^{[1]}$ can be built as follows
\begin{equation}\label{NGSvec}
X \mathcal{L}=\xi(t,q^k)\frac{\partial\mathcal{L}}{\partial t}+\eta^i(t,q^k)\frac{\partial\mathcal{L}}{\partial q^i},
\end{equation}
\begin{equation}\label{NGS1vec}
X^{[1]} \mathcal{L}=X \mathcal{L}+\dot{\eta}^k(t, q^l, \dot{q}^l)\frac{\partial\mathcal{L}}{\partial \dot{q}^k },
\end{equation}
where
\begin{equation}
\dot{\eta}^k(t, q^l,\dot{q}^l)=D_t\eta^k-\dot{q}^kD_t\xi,
\end{equation}
and
\begin{equation}\label{Dif}
D_t=\partial/\partial t+\dot{q}^k\partial/\partial q^k,
\end{equation}
is the total derivative operator. In this regard, the vector $X$
will act as a symmetry generator that constructs the conserved
quantities if there exists a gauge function, $h(t,q ^ k)$, such
that
\begin{equation}\label{condition}
X^{[1]} \mathcal{L}+\mathcal{L}(D_t\xi)=D_t \,h.
\end{equation}
The significance of Noether symmetry comes from the following
first integral of motion which asserts that if $X$ is the Noether
symmetry generator corresponding to the Lagrangian
$\mathcal{L}(t,q^i,\dot{q}^i)$, the conserved quantity associated
with this generator will be as follows
\begin{equation}\label{FirstInt}
I=-\xi\,E_{\mathcal{L}} +\eta^i\frac{\partial L}{\partial
\dot{q}^i}-h , \nonumber
\end{equation}
where $E_{\mathcal{L}}$ is the energy functional or the
Hamiltonian of the Lagrangian $\mathcal{L}$ which is defined by
\begin{equation}
E_{\mathcal{L}}\,=\,\dot{q}^i \frac{\partial \mathcal{L}}{\partial \dot{q}^i} -\mathcal{L}.
\end{equation}

As a final comment, it should be mentioned that the first integral
plays a remarkable role to obtain physically viable solutions for
the theory's system of differential equations.

\end{document}